\pdfoutput=1

\documentclass[11pt]{article}
\usepackage{amssymb}
\usepackage{amsmath}
\usepackage{amssymb}
\usepackage{amsfonts}
\usepackage{amsopn}
\usepackage{amsthm}
\usepackage{a4wide}

\usepackage{graphicx}         
\usepackage{amsfonts}
\usepackage{amssymb}
\usepackage{eucal}
\usepackage[latin1]{inputenc}
\usepackage[all]{xy}

\newtheorem{thm}{Theorem}[section]

\newtheorem{lemma}[thm]{Lemma}

\newtheorem{prop}[thm]{Proposition}
\theoremstyle{definition}
\newtheorem{defn}[thm]{Definition}
\theoremstyle{remark}

\newtheorem{remark}[thm]{Remark}

\newcommand{\e}{\epsilon}

\renewcommand{\l}{\lambda}

\newcommand{\PB}{\left\{\cdot\,,\cdot\right\}}

\newcommand{\Pb}[1]{\left\{\cdot\,,#1\right\}}
\newcommand{\pb}[1]{\left\{#1\right\}}

\renewcommand{\[}{\left[}
\renewcommand{\]}{\right]}

\newcommand{\X}{\mathcal X}

\newcommand{\Y}[3]{{\Upsilon}^{#1,#2}_{#3}}

\newcommand{\bbR}{\mathbb R}

\newcommand{\F}{\mathbf F}

\newcommand{\Z}{\mathbb Z}

\newcommand{\yt}{{\tilde y}}
\newcommand{\ds}{\displaystyle}

\newcommand{\leqs}{\leqslant}
\newcommand{\geqs}{\geqslant}
\newcommand{\pp}[2]{\frac{\partial#1}{\partial#2}}
\newcommand{\p}{\partial}
\newcommand{\we}{\wedge}

\renewcommand{\geq}{\geqs}
\renewcommand{\leq}{\leqs}

\newif\ifprivate
\privatefalse

 \numberwithin{equation}{section}

\def\???{\ifprivate {\bf {???}} \marginpar{{\Huge {\bf ?}}}\else \fi}
\numberwithin{equation}{section}

\begin{document}

\title{Integrable and superintegrable systems \\
associated with multi-sums of products}

\author{
Peter H.\ van der Kamp$^{1}$, Theodoros E. Kouloukas$^{1}$,\\
G. R. W. Quispel$^{1}$, Dinh T. Tran$^{1}$ and Pol Vanhaecke$^{2}$ \\[5mm]
$^{1}$Department of Mathematics, La Trobe University \\
Bundoora (Melbourne) VIC 3086, Australia \\
$^{2}$Laboratoire de Math\'ematiques et Applications, \\
UMR 7348 du CNRS, Universit\'e de Poitiers, \\
 86962 Futuroscope Chasseneuil Cedex, France \\[5mm]
{\em Keywords:} Integrable Systems, Lotka-Volterra system,\\
Lax representation, Superintegrability, Kahan Discretization \\[5mm]
{\em Corresponding author:} peterhvanderkamp@gmail.com
}

\maketitle
\begin{abstract}
We construct and study certain Liouville integrable, superintegrable, and non-commutative integrable systems, which are associated with multi-sums of products.
\end{abstract}

\section{Introduction}
Integrable systems have a long and distinguished history.  Starting with Newton's solution of the celestial
two-body problem, the theory of integrable ordinary differential equations (ODEs) was put on a firm footing by
Liouville. His theorem states that an autonomous Hamiltonian system in $2m$ dimensions (or equivalently $m$ degrees
of freedom), that possesses $m$ integrals in involution (i.e. whose mutual Poisson brackets all vanish) is
integrable by quadrature. In the 1960's the discovery of solitons by Zabusky and Kruskal \cite{ZK} heralded a major revival
for integrable systems. It became clear that a significent number of partial differential equations (PDEs) was also
to be regarded as integrable.  Because a PDE may be considered to be an infinite set of coupled ODEs, they actually
possess an infinite number of integrals in involution. The set of integrable PDEs includes several that have
notable applications (e.g. the Korteweg-de Vries equation, the nonlinear Schr{\"o}dinger equation and the
sine-Gordon equation). For a survey of the theory of integrable ODEs and PDEs we refer the reader to \cite{Zak,Mik}.
More recently {\it{discrete}} integrable systems have come to the fore. In these systems,
all independent variables take on discrete values. All the above mentioned integrable PDEs have discrete analogues
in the form of partial difference equations (P$\Delta$Es), which are integrable in their own right.

When imposing a periodicity condition a P$\Delta$E reduces to an O$\Delta$E or a mapping. For integrable
P$\Delta$Es the so-called staircase method yields a set of integrals for the reduced mapping \cite{QCPN,KQ}. For maps obtained
as reductions of the equations in the Adler-Bobenko-Suris (ABS) classification \cite{ABS}, for reductions of the
sine-Gordon and modified Korteweg-De Vries (mKdV) equations, and for the $p$-th order Lyness equation, first integrals were given in closed form, in terms
of multi-sums of products, $\Psi$, by using the staircase method and noncommutative Vieta expansion \cite{TKQ09}.
In particular, the Liouville integrability of mappings obtained as reductions of
the discrete sine-Gordon, mKdV, pKdV, and KdV equations was studied in detail in \cite{Tran, TKQ11, HKQT}.

This paper is concerned with integrable systems associated with a set of polynomials $z_{i}^{(n)}$, which are
related to the multisums of products $\Psi$. The polynomials and a Poisson bracket $\PB$ on $\mathbb{R}^n$
are given in Section~\ref{sec:Y_systems}. We show that
\begin{equation} \label{bz}
\{ z_{2k-1}^{(n)},z_{2l-1}^{(n)} \} = \{ z_{2k}^{(n)},z_{2l}^{(n)} \}=0
\end{equation}
for all $k, \ l \in\{1,...,\left[\frac{n}{2}\right] \}$, and therefore each polynomial
$z_{i}^{(n)}$ defines an
associated integrable Hamiltonian vector field.\footnote{Some of these results were originally obtained in the PhD thesis of Dinh Tran \cite{Tran}.}
In Section \ref{sec:subsystem}, we consider the quadratic vector
fields associated with $z_{3}^{(n)}$. This is an $n$-dimensional Lotka-Volterra system \cite{Vol,Bog}, and we prove it is superintegrable when $n$ is odd and
non-commutative integrable (of rank $2$) when $n$ is even. We also apply the Kahan discretisation (sometimes also called
Kahan-Hirota-Kimura discretisation \cite{Kah,KL,HK}) to
these quadratic vector fields, restricted to a subspace, and show that Liouville integrability and superintegrability are preserved.
\section{Integrable systems associated to multi-sums of products}\label{sec:Y_systems}
\subsection{The polynomials $z_i$ and their independence}
We introduce a set of $n$ polynomial functions $z_1,\dots,z_n$ on $\bbR^n$, and we show how they define two
integrable systems on $\bbR^n$, with respect to a (constant) Poisson structure, which will be given below.  The
polynomials $z_1,\dots,z_n$ are defined in terms of a large collection of polynomials $\Y abr$, where $a,b,r$
denote arbitrary integers, with $r\geqs0$.\footnote{These polynomials relate to polynomials $\Psi$ introduced in
\cite{TKQ09} by $\Psi_r^{a,b}=\Upsilon^{a,b+1}_{b+1-a-2r}$. They also are a special case of a much bigger class
of polynomials which has been introduced in \cite{Svi}. A different class of (non-polynomial) multi-sums of products, $\theta$, was introduced in \cite{KRQ}. For these multi-sums of products similar relations to (\ref{bz}) were derived in \cite[Lemma 1]{TKQ11}.} For $r>0$, the latter are defined in terms of $(x_i)_{i\in\Z}$ by
\begin{equation}\label{def:upsilon}
  \Upsilon^{a,b}_{r}:=\sum_{a\leq i_1<\cdots< i_r \leq b,\atop  i_j\equiv a+j-1 \hbox{\tiny{ mod }} 2} \  \prod_{j=1}^r x_{i_j}\;.
\end{equation}%
Notice that $\Y abr$ is a homogeneous polynomial of degree $r$ and that it depends only on the variables
$x_a,x_{a+1},\dots,x_b$; in particular,
\begin{equation}\label{eq:Y_0}
  \Y abr=0\;\,\hbox{ when } b-a+1<r\;.
\end{equation}%
Moreover, $\Y abr$ satisfies, for all $a,b$ and for $r>1$, the following two recursion relations
\begin{equation}\label{eq:Y_rec}
  \Y abr=\Y{a+2}br+x_a\Y{a+1}b{r-1}\;,\qquad\Y abr=\Y a{b-1}r+\epsilon^{b-a}_rx_b\Y a{b-1}{r-1}\;,
\end{equation}%
where
\begin{equation*}
  \epsilon^q_r:=\left\{
  \begin{array}{cl}
    0&\hbox{ if $q$ and $r$ have the same parity},\\
    1&\hbox{ otherwise}.
  \end{array}
  \right.
\end{equation*}%
In order for these recursion relations to make sense and be correct also for $r=1$, the following additional
definition is needed:
\begin{equation}\label{eq:initial_cond}
  \Y ab0:=\left\{
  \begin{array}{ccl}
    0&\hbox{ if }&a-1>b\;,\\
    1&\hbox{ if }&a-1\leqs b\;.
  \end{array}
  \right.
\end{equation}%
It is easy to see that the functions $\Y abr$ are uniquely determined by either one of the recursion relations in
(\ref{eq:Y_rec}), together with (\ref{eq:Y_0}) and (\ref{eq:initial_cond}). In terms of the functions~$\Y abr$, we
define
\begin{equation}\label{eq:z}
z_i:=\Y 1ni
\end{equation}
for $i=1,\dots,n$ and we view each $z_i$ as a polynomial function on $\bbR^n$; when~$n$ is not
clear from the context, we also write $z_i^{(n)}$ for $z_i$. It is also convenient to extend the definition
(\ref{eq:z}) to arbitrary $i\geqs0$, by defining $z_0:=1$ and $z_i:=0$ for $i>n$. In terms of the functions $z_i$, the
second recursion relation in (\ref{eq:Y_rec}) leads to
\begin{equation}\label{eq:z_rec}
    z_i^{(n+1)}=z_i^{(n)}+\epsilon_i^nx_{n+1}z_{i-1}^{(n)}\;,
\end{equation}%
for any $i>0$, i.e., $z_i^{(n+1)}=z_i^{(n)}$ if $i$ and $n$ have the same parity and
$z_i^{(n+1)}=z_i^{(n)}+x_{n+1}z_{i-1}^{(n)}$ otherwise. In particular, $z_i^{(n)}={z_i^{(n+1)}}_{\big\vert
x_{n+1}=0}$ for all $i$ and $n$. Also, each polynomial $z_i$ is homogeneous and has degree $i$. Here are a few
low-dimensional examples:
\begin{eqnarray*}
  \hbox{dimension } 1&:& z_1=x_1\;,\\
  \hbox{dimension } 2&:& z_1=x_1\;,\ z_2=x_1x_2\;,\\
  \hbox{dimension } 3&:& z_1=x_1+x_3\;,\ z_2=x_1x_2\;,\ z_3=x_1x_2x_3\;,\\
  \hbox{dimension } 4&:& z_1=x_1+x_3\;,\ z_2=x_1x_2+x_1x_4+x_3x_4\;,\ z_3=x_1x_2x_3\;,\ z_4=x_1x_2x_3x_4\;,\\
  \hbox{dimension } 5&:& z_1=x_1+x_3+x_5\;,\ z_2=x_1x_2+x_1x_4+x_3x_4\; \ z_3=x_1x_2x_3+x_1x_2x_5+x_1x_4x_5\\&&+x_3x_4x_5\;,\ z_4=x_1x_2x_3x_4\;, z_5=x_1x_2x_3x_4x_5\;.
\end{eqnarray*}
\begin{prop}\label{prp:independent}
  On the open dense subset $D$ of $\bbR^n$, defined by 
  \begin{equation}\label{C7E:domain}
    D:=\{(x_1,x_2,\ldots,x_n)\in \bbR^n\mid x_1x_2\ldots x_{n-1}\neq 0\}
  \end{equation}
  the functions $z_1,\dots,z_n$ have independent differentials, hence they define a coordinate system on a
  neighborhood of any point of $D$.
\end{prop}
\begin{proof}
We need to show that at every point of $D$ the rank of the Jacobian matrix
\begin{equation}\label{C7E:Jaco}
  J^{(n)}:=\displaystyle{\frac{\partial(z_1,\dots,z_n)}{\partial (x_1,x_2,\ldots, x_{n})}}=\left(\pp{z_i^{(n)}}{x_j}\right)_{ij}
\end{equation}
is equal to $n$.  To do this, we use an LU-decomposition of $J^{(n)}$, i.e. we write $J^{(n)}=L^{(n)}U^{(n)}$,
where~$L^{(n)}$ is a lower triangular matrix and $U^{(n)}$ is an upper triangular matrix; we show that all entries
on the diagonal of $L^{(n)}$ and of $U^{(n)}$ are non-zero at every point of $D$. Precisely, we show that the upper
triangular entries of $L^{(n)}$ are given by $L^{(n)}_{ij}=\Upsilon^{j+1,n}_{i-j}$, so that all diagonal entries of
$L^{(n)}$ are equal to $1$, and that the diagonal entries of $U^{(n)}$ are given by $U^{(n)}_{kk}=x_1x_2\dots
x_{k-1}$. We do this by induction. For $n=1$ it is clear, so let us assume that $J^{(n)}=L^{(n)}U^{(n)}$ for some
$n>0$, with $L^{(n)}$ and $U^{(n)}$ as above. As before, we usually drop the superscript~$n$ and simply write
$J=LU$. We need to prove that we can write $J^{(n+1)}=L^{(n+1)}U^{(n+1)}$, with $L^{(n+1)}$ lower triangular,
$U^{(n+1)}$ upper triangular, with entries as given above. Using the recursion relation (\ref{eq:z_rec}), we can
write $J^{(n+1)}$ in terms of~$J$, to wit,
\begin{equation}\label{C7E:sum}
  J^{(n+1)}=\begin{pmatrix}J&\bf{0}\\ \bf{0}&0\end{pmatrix}+
            \begin{pmatrix}\bf{0}&0\\  \bar{J}&\bf{0}\end{pmatrix}+
            \begin{pmatrix}
              0&0&\ldots & 0 & \epsilon^{n}_1\\
              0&0&\ldots&0&    \epsilon^{n}_2z_1\\
              0&0&\ldots &0&   \epsilon^{n}_3z_2\\
              \vdots&\vdots&\ddots &\vdots&\vdots\\
              0&0&\ldots& 0&   \epsilon^{n}_{n+1}z_{n}
            \end{pmatrix}\;,
\end{equation}
where $\bar{J}$ is the $(n\times n)$-matrix obtained from $J$ by multiplying its $k^{th}$ row by
$\epsilon^n_{k+1}x_{n+1}$, for $k=1,\dots,n$.  We similarly define $\bar L$, starting from $L$. Then the relation
$J=LU$ implies $\bar J=\bar LU$, so that the first two terms in (\ref{C7E:sum}) can be written as 
\begin{equation}
  \left(
  \begin{pmatrix}L&\bf{0}\\ \bf{0}&0\end{pmatrix}+
  \begin{pmatrix}\bf{0}&0\\ \bar{L}&\bf{0}\end{pmatrix}
  \right)
  \begin{pmatrix}U&\bf{0}\\ \bf{0}&0\end{pmatrix}=
  L^{(n+1)}
  \begin{pmatrix}U&\bf{0}\\ \bf{0}&0\end{pmatrix},
\end{equation}
where $L^{(n+1)}$ can be chosen as a lower triangular matrix: $L$ itself is lower triangular, with $1's$ on the
diagonal, and we choose the last diagonal entry of $L^{(n+1)}$ also equal to $1$. For $1\leqs j< i\leqs n$ we have,
using (\ref{eq:Y_rec}) and the induction hypothesis,
\begin{equation}
  L^{(n+1)}_{ij}=L_{ij}+\bar{L}_{i-1,j}=\Upsilon^{j+1,n}_{i-j}+\epsilon^n_ix_{n+1}\Upsilon^{j+1,n}_{i-j-1}=\Upsilon^{j+1,n+1}_{i-j}\;;
\end{equation}
similarly, if $i=n+1$ and $j<n$, then
$$
  L^{(n+1)}_{n+1,j}=\bar{L}_{nj}=x_{n+1}\Y{j+1}n{n-j}=\Y{j+1}{n+1}{n-j+1}\;.
$$
This shows that $L^{(n+1)}$ and its entries have the asserted form. We now turn our attention to~$U^{(n+1)}$. Let
us denote the last column of the last term in~(\ref{C7E:sum}) by $\bf{j}$, so ${\bf j}_i=\epsilon^n_iz_{i-1}$, for
$i=1,\dots,n+1$.  Since $L^{(n+1)}$ is invertible there exists a unique column vector $\bf u$ such that
$L^{(n+1)}{\bf u}={\bf j}$.  It leads to the LU decomposition $J^{(n+1)}=L^{(n+1)}U^{(n+1)}$, where $U^{(n+1)}$ is
the upper triangular matrix, defined by
$$
  U^{(n+1)}:=\begin{pmatrix}
    U&\bf{0}\\
    \bf{0}&0
  \end{pmatrix}+
  \begin{pmatrix}
    \bf{0}&{\bf u}
  \end{pmatrix}.
$$
In order to prove that the entries of $U^{(n+1)}$ have the asserted form, it suffices to show that ${\bf
u}_{n+1}=x_1x_2\dots x_n$. If we denote the last row of $(L^{(n+1)})^{-1}$ by~$\bf t$, then ${\bf u}_{n+1}={\bf
tj}$; we will show that ${\bf t}=(0,\dots,0,-x_{n+1},1)$, from which we obtain
\begin{equation*}
  {\bf u}_{n+1}={\bf tj}=-\epsilon_n^nx_{n+1}z_{n-1}+\epsilon_{n+1}^nz_n=z_n=x_1x_2\dots x_n\;,
\end{equation*}%
as was to be shown. In order to prove the proposed formula for $\bf t$ it suffices to show that
$(0,\dots,0,-x_{n+1},1)L^{(n+1)}=(0,0,\dots,0,1)$, which is tantamount to saying that $L^{(n+1)}_{n+1,n+1}=1$
(which is true by definition) and that $L_{n+1,k}^{(n+1)}-x_{n+1}L_{n,k}^{(n+1)}=0$ for $k=1,\dots,n$. In terms of
the matrices $L$ and $\bar L$ this amounts to $\bar L_{nk}=x_{n+1}L_{nk}$, which is precisely the definition of the
last row of $\bar L$.%
\end{proof}
\begin{remark}
The map $x\to z$ is in fact a birational map, i.e., one can write $x_1,\dots,x_n$ as rational functions of
$z_1,\dots,z_n$ (while each $z_i$ is a polynomial in the $x_k$). The proof goes by induction on $n$. Suppose that
we have shown that $x_1,\dots,x_n$ can be written as rational functions of $z_1^{(1)},\dots,z_n^{(n)}$,
\begin{equation*}
  x_k=R_k(z_1^{(n)},\dots,z_n^{(n)})=R_k\left(z_i^{(n)}\right)_{i=1,\dots,n}\;,
\end{equation*}%
where the latter notation will come in handy soon.  Repeated use of (\ref{eq:z_rec}) leads to
\begin{eqnarray*}
  z_{n+1}^{(n+1)}&=&x_{n+1}z_n^{(n)}=\dots=x_1x_2\dots x_{n+1}\;,\\
  z_{n}^{(n+1)}&=&z_n^{(n)}=x_1x_2\dots x_{n}\;,
\end{eqnarray*}
so that 
\begin{equation}\label{eq:rat_last}
  x_{n+1}=z_{n+1}^{(n+1)}/z_n^{(n+1)}\;.
\end{equation}%
Substituted in (\ref{eq:z_rec}) we get, for $i=1,\dots,n$,  
\begin{equation*}
  z_i^{(n)}=z_i^{(n+1)}-\epsilon_i^n \frac{z_{n+1}^{(n+1)}}{z_n^{(n+1)}} z_{i-1}^{(n)}
           =z_i^{(n+1)}-\epsilon_i^n \frac{z_{n+1}^{(n+1)}}{z_n^{(n+1)}} z_{i-1}^{(n+1)}\;,
\end{equation*}%
where we used in the last step that if $\epsilon_i^n\neq 0$ (so that $n$ and $i$ have opposite parity), then
$z_{i-1}^{(n+1)}=z_{i-1}^{(n)}$. It follows that
\begin{equation}\label{eq:rat_first}
  x_k=R_k\left(z_i^{(n+1)}-\epsilon_i^n \frac{z_{n+1}^{(n+1)}}{z_n^{(n+1)}} z_{i-1}^{(n+1)}\right)_{i=1,\dots,n}\;,
\end{equation}%
for $k=1,\dots,n$, Equations (\ref{eq:rat_last}) and (\ref{eq:rat_first}), together, show that $x_1,\dots,x_{n+1}$
can be written as rational functions of $z_1^{(1)},\dots,z_{n+1}^{(n+1)}$.  Combined with
Proposition~\ref{prp:independent} it shows that $z_1,\dots,z_n$ form a coordinate system on $D$.
\end{remark}

\subsection{The Poisson structure and involutivity}
On $\bbR^n$ we consider the constant Poisson structure, defined by
\begin{equation} \label{PB1}
  \pb{x_i,x_j}_n:=\delta_{i+1,j}-\delta_{i,j+1}\;.
\end{equation}
When $n$ is even, its rank is $n$; otherwise its rank is $n-1$ and $z_1=\sum_{i=1}^{(n+1)/2}x_{2i-1}$ is a Casimir
function, since $\pb{x_i,z_1}=0$ for all $i$. In geometrical terms, the Poisson structure in the odd-dimensional
case is obtained by a Poisson reduction (see \cite[Ch.\ 5.2]{LPV}) from the Poisson structure in the
even-dimensional case: if we view $\bbR^{2m-1}$ as the quotient of $\bbR^{2m}$ under the quotient map
$\pi:\bbR^{2m}\to \bbR^{2m-1}$, defined by $(x_1,x_2,\dots,x_{2m})\mapsto (x_1,x_2,\dots,x_{2m-1})$, then the pair
$(\bbR^{2m},\bbR^{2m-1})$ is Poisson reducible and the reduced Poisson structure, inherited from $\PB_{2m}$ is
$\PB_{2m-1}$. In particular, $\pi:(\bbR^{2m},\PB_{2m})\to(\bbR^{2m-1},\PB_{2m-1})$ is a Poisson map.

In the following proposition we give explicit formulas for the Poisson brackets between the functions
$z_1,\dots,z_n$. We write here, and in the rest of the paper, $a\equiv b$ when $a\equiv b\mod2$, i.e., when the
integers $a$ and $b$ have the same parity.
\begin{prop}\label{prp:z_brackets}
  For any $i$ and $j$ with $0\leqs i, j\leqs n$, we have that
\begin{equation*}
  \pb{z_i,z_j}_n=\left\{
  \begin{array}{ccl}
    0&\hbox{ if }&i\equiv j\;,\\
    \sum\limits_{\tiny\begin{array}{c}k+\ell=i+j\\ 0<k<i\end{array}}(-1)^{i-k}z_{k-1}z_{l-1}&\hbox{ if }&i\equiv j+1\equiv n\;.
  \end{array}
  \right.
\end{equation*}%
\end{prop}
\begin{proof}
The proof proceeds by induction on $n$.  It is easy to check that the formulas hold for $n=1$. Suppose that they
hold for some $n\geqs 1$. We need to prove that they hold for $n+1$. Let $0\leqs i,j\leqs n+1$ be arbitrary
integers. If $i=0$ or $j=0$ the formula is easily checked (recall that $z_0=1$), so let us assume that $i,j>0$.
Using~\eqref{eq:z_rec}, we have
\begin{equation}\label{E:PoiIndu}
  \pb{z_i^{(n+1)},z_j^{(n+1)}}_{n+1}=\pb{z^{(n)}_i+\epsilon^{n}_i x_{n+1}z^{(n)}_{i-1},z^{(n)}_j+\epsilon^{n}_jx_{n+1}z^{(n)}_{j-1}}_{n+1}\;.
\end{equation}
We use $\dot F$ as a shorthand for $\pb{F,x_{n+1}}_{n+1}=\p F/\p{x_{n}}$ and we write $z_k$ for $z_k^{(n)}$ for any
$k$. We have that $\pb{z_i,z_j}_{n+1}=\pb{z_i,z_j}_{n}$ because $z_i$ and $z_j$ depend on $x_1,\dots,x_n$ only.  We
distinguish 3 cases, according to the relative parity of $i,j$ and~$n$.

\noindent $\bullet$ $\underline{i\equiv j\equiv n}\quad$ Then $\epsilon^{n}_i=\epsilon^{n}_j=0$ and $j<n+1$, so
that (\ref{E:PoiIndu}) becomes
$$
  \pb{z_i^{(n+1)},z_j^{(n+1)}}_{n+1}=\pb{z_i,z_j}_{n}=0\;.
$$
\noindent $\bullet$ $\underline{i\equiv j\equiv n+1}\quad$ Then $\epsilon^{n}_i=\epsilon^{n}_j=1$. We expand the
right hand side of~(\ref{E:PoiIndu}) and use that $\pb{z_i,z_j}=0$ and that $\dot z_i=\dot z_j=0$ ($z_i$ and $z_j$
are independent of $x_n$ since $i\equiv j\equiv n+1$) to obtain
\begin{equation*}
  \pb{z_i^{(n+1)},z_j^{(n+1)}}_{n+1} =x_{n+1}(\pb{z_i,z_{j-1}}_n+\dot z_{i-1}z_{j-1}-\pb{z_j,z_{i-1}}_n-\dot z_{j-1}z_{i-1})\;.
\end{equation*}%
Since $i$ and $n$ have opposite parity, we find from (\ref{eq:z_rec}) that
$$
  \dot z_{i-1}=\frac{\p z_{i-1}^{(n)}}{\p x_n}=z_{i-2}^{(n-1)}=z_{i-2}^{(n)}=z_{i-2}\;,
$$
and similarly for $\dot z_{j-1}$, so it suffices to show that 
\begin{equation}\label{eq:comm}
  \pb{z_{i-1},z_{j}}_n+ z_{i-2}z_{j-1}-\pb{z_{j-1},z_{i}}_n- z_{j-2}z_{i-1}=0\;.
\end{equation}%
Using the  induction hypothesis, the left hand side in (\ref{eq:comm}) is given by
$$
  \sum_{\tiny\begin{array}{c}k+\ell=i+j-1\\ 0<k<i\end{array}}(-1)^{i-1-k}z_{k-1}z_{\ell-1}-
  \sum_{\tiny\begin{array}{c}k+\ell=i+j-1\\ 0<k<j\end{array}}(-1)^{j-1-k}z_{k-1}z_{\ell-1}
$$
which is zero, because every term appears twice with opposite signs (recall that $i\equiv j$).

\noindent $\bullet$ $\underline{i\equiv j+1}\quad$ By interchanging $i$ and $j$ if needed, we may suppose that
$j\equiv n$. Then $\epsilon^{n}_i=1$ and $\epsilon^{n}_j=0$ so that we get, as above,
$$
  \pb{z_i^{(n+1)},z_j^{(n+1)}}_{n+1}=\pb{z_i,z_j}_n-z_{i-1}\dot z_j
  =\sum_{\tiny\begin{array}{c}k+\ell=i+j\\ 0<k\leqs j\end{array}}(-1)^{j-k-1}z_{k-1}z_{\ell-1}\;.
$$
This is to be compared with
\begin{eqnarray*}
  \lefteqn{\sum\limits_{\tiny\begin{array}{c}k+\ell=i+j\\ 0<k<i\end{array}}(-1)^{i-k}z_{k-1}^{(n+1)}z_{\ell-1}^{(n+1)}}\\
  &=&\sum\limits_{\tiny\begin{array}{c}k+\ell=i+j\\ 0<k<i\end{array}}(-1)^{i-k}
     \left(z_{k-1}+\epsilon^{n}_{k-1}x_{n+1}z_{k-2}\right)\left(z_{\ell-1}+\epsilon^{n}_{\ell-1}x_{n+1}z_{\ell-2}\right)\\
  &=&\sum\limits_{\tiny\begin{array}{c}k+\ell=i+j\\ 0<k<i\end{array}}(-1)^{i-k}z_{k-1}z_{\ell-1}
  + x_{n+1}\left[\sum_{\tiny\begin{array}{c}k+\ell=i+j\\ 0<k<i-1\\ k\equiv i\end{array}}z_{k-1}z_{\ell-2}
   -\sum_{\tiny\begin{array}{c}k+\ell=i+j\\ 1<k<i\\ k\equiv i+1\end{array}}z_{k-2}z_{\ell-1}\right]\\
  &=&\sum\limits_{\tiny\begin{array}{c}k+\ell=i+j\\ 0<k<i\end{array}}(-1)^{i-k}z_{k-1}z_{\ell-1}\;.
\end{eqnarray*}
Taking the difference of both expressions we get zero, again because in the difference every term appears twice
with opposite signs.
\end{proof}
\subsection{The $\Upsilon$-systems, their Liouville integrability and linearisation}\label{parg:integrability}
Recall the definition of a Liouville integrable system:
\begin{defn}
On an ($n=2r+s$)-dimensional Poisson manifold where the Poisson bracket has rank $2r$, a tuple of $n-r=r+s$ functionally
independent functions is {\em Liouville integrable} if they are (pairwise) in involution.
\end{defn}
According to Proposition \ref{prp:z_brackets}, the functions $z_i$ with even (resp.\ odd) index $i$ are pairwise in
involution. Since for $n$ odd the function $z_1$ is a Casimir, hence is in involution with all functions~$z_i$, we
define
$$
  \begin{array}{rclrcll}
    \F&:=&(z_1,z_3,\dots,z_{n-1})\;,\quad &\F'&:=&(z_2,z_4,\dots,z_{n})\quad&\hbox{if $n$ is even},\\
    \F&:=&(z_1,z_3,\dots,z_n)\;,\quad &\F'&:=&(z_1,z_2,z_4,\dots,z_{n-1})\quad&\hbox{if $n$ is odd}.
\end{array}
$$
We use the above results to show in the following theorem that both $\F$ and $\F'$ are integrable systems on
$(\bbR^n,\PB_n)$.
\begin{thm}\label{thm:Y_invol}
  Both $\F$ and $\F'$ are Liouville integrable systems on $(\bbR^n,\PB_n)$. On the open subset $D$, defined in
  Proposition \ref{prp:independent}, the functions $z_1,\dots,z_n$ define a coordinate system in terms of which all
  the Hamiltonian vector fields $\X_{z_i}$ are linear.
\end{thm}
\begin{proof}
Recall that the rank of the Poisson structure $\PB_n$ is equal to $n$ when $n$ is even, and is otherwise equal to
$n-1$. When $n$ is even both $\F$ and $\F'$ contain $n/2$ functions which are independent (according to
Proposition~\ref{prp:independent}) and are pairwise in involution (Proposition \ref{prp:z_brackets}). Thus, both
$\F$ and $\F'$ define Liouville integrable systems on $(\bbR^n,\PB_n)$. When $n$ is odd the rank of the Poisson
structure is $n-1$ and so we need, besides the Casimir $z_1$, another $(n-1)/2$ independent functions in
involution. Using Propositions \ref{prp:independent} and \ref{prp:z_brackets} we can again conclude that both $\F$
and $\F'$ define Liouville integrable systems on $(\bbR^n,\PB_n)$.

In terms of the coordinates $z_1,\dots,z_n$ on $D$, the Hamiltonian vector fields $\X_{z_i}:=\Pb{z_i}_n$ take a particularly
simple form. We show this for even $n$. According to Proposition \ref{prp:z_brackets}, these vector fields are
given by:
\begin{eqnarray*}
  \X_{z_{2s}}&:&\left\{
    \begin{array}{rcl}\dot z_{2r}&=&0\;,\\ \dot z_{2r-1}&=&\sum_{i=1}^{2s-1}(-1)^{i-1}z_{2s-i-1}z_{2r+i-2}\;,\end{array}
    \right.\\
  \X_{z_{2s-1}}&:&\left\{
    \begin{array}{rcl}\dot z_{2r-1}&=&0\;,\\ \dot z_{2r}&=&\sum_{i=1}^{2r-1}(-1)^iz_{2r-i-1}z_{2s+i-2}\;,\end{array}
    \right.
\end{eqnarray*}
where the dot denotes differentiation, i.e. $\dot{z}=\frac{\text{d}z}{\text{d}t}$.
Each one of these vector fields becomes a linear vector field after the first group of equations has been
integrated (giving $z_{2r}=c_{2r}$ for $\X_{z_{2s}}$ and $z_{2r-1}=c_{2r-1}$ for $\X_{z_{2s-1}}$, where the $c_i$
are constants).
\end{proof}
In the sequel, we refer to the integrable systems $(\bbR^n,\PB_n,\F)$ (resp.\ the integrable systems
$(\bbR^n,\PB_n,\F')$) as the \emph{odd} (resp.\ the \emph{even}) \emph{$\Upsilon$-systems} in dimension $n$.

\subsection{Lax equations for the $\Upsilon$-systems}\label{parg:lax}
In this subsection, we show that the integrable vector fields $\X_{z_i}$ of the $\Upsilon$-systems, defined in
Section \ref{sec:Y_systems}\ref{parg:integrability}, are given by Lax equations, i.e. for each integrable system
(in both odd and even dimensions) we provide a matrix $L$ and matrices $B_i$, such that each vector field of
the system acts on $L$ as the commutator with $B_i$;
$$
\X_{z_{i}}(L)=[L,B_i]\;.
$$
This yields an alternative proof (i.e., without using Proposition \ref{prp:z_brackets}) that the functions $z_{i}$ and $z_{j}$ are in
involution, when $i\equiv j$.  We first consider the case of the even $\Upsilon$-systems in the even-dimensional
case ($n$ even) and derive from it the odd-dimensional case ($n$ odd); for the case of the odd $\Upsilon$-system we
first treat the odd-dimensional case and derive then the even-dimensional case from it.
 
\underline{The case of $\F'$ with $n$ even.} We set $n=2m$. We show that a Lax operator for $\F'$ on $\bbR^n$ is
given by the following $n\times n$ matrix:
\begin{equation}\label{eq:lax_n_even_z_even}
  L'_{2m}=
  \begin{pmatrix}%
    0&x_1&0&x_1&0&x_1&\cdots&0&x_1\\
    -x_2&0&0&0&0&0&\cdots&0&0\\
    0&0&0&x_3&0&x_3&\cdots&0&x_3\\
    -x_4&0&-x_4&0&0&0&\cdots&0&0\\
    \vdots&\vdots&\vdots&\vdots&\vdots&\vdots&&\vdots&\vdots\\
    0&0&0&0&0&0&\cdots&0&x_{2m-1}\\
    -x_{2m}&0&-x_{2m}&0&-x_{2m}&0&\cdots&-x_{2m}&0
  \end{pmatrix}\;.
\end{equation}
%
%
To do this, we first show that the polynomials in $\F'=(z_2,z_4,\dots,z_{2m})$ appear as coefficients of the
characteristic polynomial of $L'_{2m}$.  First notice that it follows at once from a Laplace expansion of the
determinant with respect to the last two rows that $\det(L'_{2m})=x_1x_2\dots x_n$. For any $s$, we define a
tridiagonal matrix $M'_s$ by setting $(M'_s)_{i,j}:=x_j^{-1}(\delta_{i,j+1}-\delta_{j,i+1})$. It is easy to verify
that $M'_{2m}$ is the inverse of $L'_{2m}$. Thus,
\begin{equation}\label{eq:from_L_to_M}
  \vert L'_{2m}-\lambda I_{2m}|=\vert L'_{2m}\vert\;\vert I_{2m}-\l M'_{2m}\vert=\l^{2m}\vert L'_{2m}\vert\;\vert
  M'_{2m}-\frac1\l I_{2m}\vert\;,
\end{equation}%
and it suffices to compute the characteristic polynomial of $M'_{2m}$. We claim that for any $s$, 
\begin{equation}\label{eq:p_n_rec}
  p'_s(\mu):=\vert M'_s-\mu I_s\vert= \frac{(-1)^s}{z_s^{(s)}}\sum_{k=0}^{\[s/2\]}z_{s-2k}^{(s)}\mu^{s-2k}\;.
\end{equation}
Notice that the formula for $p_s'$ is obviously correct for $s=1$ and $s=2$. Expanding $\vert M'_s-\mu I_s\vert$ along
its last column, one finds that
\begin{equation}\label{eq:pp_rec}
  p'_{s+2}(\mu)=\frac{p'_s(\mu)}{x_{s+1}x_{s+2}}-\mu p'_{s+1}(\mu)\,
\end{equation}
for all $s\geqs1$, so it suffices to show that the formula for $p'_s(\mu)$, given in (\ref{eq:p_n_rec}) satisfies
the recursion relation (\ref{eq:pp_rec}). That this is indeed so follows easily from (\ref{eq:z_rec}), combined with
the formula $z_s^{(s)}=x_1x_2\dots x_s$. Combined with (\ref{eq:from_L_to_M}), this proves that the characteristic
polynomial of $L'_{2m}$ is given by
\begin{equation*}
  \vert L'_{2m}-\lambda I_{2m}|=\sum_{i=0}^m z_{2i}\l^{2m-2i}\;;
\end{equation*}%
in particular, its coefficients are precisely the Hamiltonians $z_2,z_4,\dots,z_n$ which make up $\F'$. For
$k=1,\dots,m$, a matrix $B_{2m,2k}'$ satisfying
\begin{equation}\label{E:Lax_even_even}
  \X_{z_{2k}}(L'_{2m})=[L'_{2m},B_{2m,2k}']\;,
\end{equation}%
can be chosen upper triangular and with entries $b_{ij}$ given by
\begin{equation}\label{eq:B2m2k}
  b_{ij}:=
  \left\{
  \begin{array}{ll}
    \ds \sum_{\substack{0\leqs r\leqs 2k-2 \\ r\equiv i}}(\Y1ir\delta_{ij}-x_i
    \ds \Y1{i-1}{r-1})\Y{j+1}n{2k-2-r}\qquad&\hbox{if $i\leqs j$ and $i\equiv j$}\\
    \qquad0&\hbox{otherwise.}
  \end{array}
  \right.
\end{equation}%
We will give a  proof that the matrices $L'_{2m}$ and $B'_{2m,2k}$ satisfy \eqref{E:Lax_even_even} in Appendix
\ref{AC}.  

\underline{The case of $\F'$ with $n$ odd.} We set $n=2m-1$ and construct a Lax operator for $\F'$ (in dimension
$2m-1$), by slightly modifying the matrix $L'_{2m}$ (which has size $2m$): we substitute $0$ for $x_{2m}$ in
$L'_{2m}$ (making all entries on its last row equal to zero) and for all entries in the last column except the
second-to-last entry (which is equal to $x_{2m-1}$). This yields a Lax operator $L'_{2m-1}$ for $\F'$ in the odd
dimension $2m-1$. Its characteristic polynomial is given by
\begin{equation}\label{eq:one_down}
  \vert L'_{2m-1}-\l I_{2m}\vert=\l^2\vert L'_{2m-2}-\l I_{2m}\vert=\sum_{i=0}^{m-1}z_{2i}^{(2m-2)}\l^{2m-2i} 
  =\sum_{i=0}^{m-1}z_{2i}^{(2m-1)}\l^{2m-2i}\;, 
\end{equation}%
where we used that $z_{2i}^{(2m-2)}=z_{2i}^{(2m-1)}$, an immediate consequence of (\ref{eq:z_rec}). This shows
that, except for the Casimir $z_1$, all functions in $\F'=(z_1,z_2,z_4,\dots,z_{n-1})$ appear as coefficients of the
characteristic polynomial of $L'_{2m-1}$. For $k=1,\dots,m-1$, the matrix $B'_{2m-1,2k}$ is obtained from
$B'_{2m,2k}$ as follows: replace all entries in its last two rows and in its last two columns by $0$, except for
the entry at position $(2m-1,2m-1)$, which is set equal to $z_{2k-1}^{(2m-3)}/x_{2m-1}$. For a proof that the
matrices $L'_{2m-1}$ and $B'_{2m-1,2k}$ satisfy a Lax equation as in \eqref{E:Lax_even_even}, we refer again to
Appendix~\ref{AC}.
%

\underline{The case of $\F$ with $n$ odd.} We set, as before, $n=2m-1$. A Lax operator for $\F=(z_1,z_3,\dots,z_n)$
on $\bbR^n$ is given by the following $n\times n$ matrix:
\begin{equation}\label{eq:lax_n_odd_z_odd}
  L_{2m-1}=
  \begin{pmatrix}%
    x_1&x_1&x_1&x_1&x_1&\cdots&x_1&x_1\\
    0&0&x_2&0&x_2&\cdots&0&x_2\\
    x_3&0&x_3&x_3&x_3&\cdots&x_3&x_3\\
    0&0&0&0&x_4&\cdots&0&x_4\\
    x_5&0&x_5&0&x_5&\cdots&x_5&x_5\\
    \vdots&\vdots&\vdots&\vdots&\vdots&&\vdots&\vdots\\
    x_{2m-1}&0&x_{2m-1}&0&x_{2m-1}&\cdots&0&x_{2m-1}
  \end{pmatrix}
\end{equation}
It can be shown by induction that $\det L_{2m-1}=x_1x_2\dots x_{2m-1}$. For any $s$, consider the matrix $M_s$,
which is obtained from the matrix $M'_s$ by replacing the entry in its upper right corner (which is zero) by
$x_{2m-1}^{-1}.$ It is easy to verify that $M_{2m-1}$ is the inverse of $L_{2m-1}$. Expanding $\vert M_s-\mu
I_s\vert$ along its last column, we find that $p_s(\mu)=p_s'(\mu)+(x_1x_2\dots x_s)^{-1}$. Using the explicit
formula~(\ref{eq:p_n_rec}) for $p_s'(\mu)$ it follows, as in (\ref{eq:from_L_to_M}), that
\begin{equation*}
  \vert L_{2m-1}-\lambda I_{2m-1}|=-\l^{2m-1}+\sum_{i=1}^{m} z_{2i-1}\l^{2m-2i}\;;
\end{equation*}%
in particular, its coefficients are precisely the Hamiltonians $z_1,z_3,\dots,z_n$ which make up $\F$. The matrix
$B_{2m-1,2k-1}$ is defined as in (\ref{eq:B2m2k}), but with $2k$ replaced by $2k-1$. Notice that this yields that
$B_{2m-1,1}=0$, which is correct since $z_1$ is a Casimir. The proof that 
\begin{equation*}
  \X_{z_{2k-1}}(L_{2m-1})=[L_{2m-1},B_{2m-1,2k-1}]\;,
\end{equation*}%
for $k=2,\dots,m$ is an easy adaptation of the proof given in Appendix \ref{AC}. 

\underline{The case of $\F$ with $n$ even.}  We set $n=2m$ and we construct the Lax operator $L_{2m}$ from
$L_{2m+1}$ by substituting in it $0$ for $x_{2m+1}$ (making all entries on its last row equal to zero) and for
all entries in the last column except the second-to-last entry (which is equal to $x_{2m}$). One obtains, as in
(\ref{eq:one_down}),
\begin{equation*}
  \vert L_{2m}-\l I_{2m+1}\vert=\l^2\vert L_{2m-1}-\l I_{2m-1}\vert=-\l^{2m+1}+\sum_{i=1}^{m}z_{2i-1}^{(2m)}\l^{2m+2-2i}\;. 
\end{equation*}%
The matrix $B_{2m,2k-1}$ is obtained from the matrix $B_{2m+1,2k-1}$ as follows: replace all entries in its last
two rows and in its last two columns by $0$, except for the following entries
$$
  b_{1,2m}=-z^{(2m)}_{2k-2}/x_{2m},\quad  b_{2m,2m}=z^{(2m)}_{2k-2}/x_{2m},\quad b_{2,2m+1}=-z^{(2m)}_{2k-2}/x_1\;.
$$
More details can be found in Appendix C.        
\section{On the quadratic vector fields}\label{sec:subsystem}
The integrable vector fields of the $\Upsilon$-systems are homogeneous, namely the Hamiltonian vector field
associated to $z_i$ is homogeneous of degree $i-1$, since $z_i$ is homogeneous of degree $i$ (recall that the
Poisson structure is constant). In this section we study the quadratic vector fields of the $n$-dimensional
$\Upsilon$-systems. We show that they are closely related to a class of Lotka-Volterra systems in dimension $m$,
where $m=[(n+1)/2]$. We establish the super- and Liouville integrability of these Lotka-Volterra systems, by exhibiting explicit rational constants of motion; notice that the only non-trivial dimension in which the
integrability of these Lotka-Volterra systems was known is in dimension~4, see \cite[Example 12]{Dam}. We derive from it the superintegrability (if the dimension $n$ is odd) and non-commutative integrability of rank 2 (if the dimension $n$ is even) of the quadratic vector fields; recall that their Liouville integrability was obtained in the previous section. For the reader who is not familiar with these notions, here are the definitions.
\begin{defn}
A vector field on a $n$-dimensional manifold is {\em superintegrable} if it admits $n-1$ functionally independent constants of motion.
\end{defn}
In particular, a Hamiltonian vector field is superintegrable if there is a functionally independent set of $n-1$ functions,
including the Hamiltonian, which are in involution with the Hamiltonian.
\begin{defn}
On an $n=2r+s$-dimensional Poisson manifold $M$ where the Poisson bracket has rank $\geq 2r$, a tuple of $n-r=r+s$ functionally
independent functions is {\em non-commutative integrable of rank $r$}, if $r$ functions are in involution with all $n-r$
functions, and their Hamiltonian vector fields are linearly independent at some point of $M$.
\end{defn}
Thus, on a Poisson manifold, superintegrability is equivalent to non-commutative integrability of rank 1, and Liouville
integrability is equivalent to non-commutative integrability of rank $r$, although here we would rather say that the system
is commutative integrable. In both cases the rank of the system is at most half the rank of the Poisson bracket, with equality in the commutative (Liouville) case.

At the end of the section, we also derive from the superintegrability of the special Lotka-Volterra subsystems the
superintegrability of their Kahan discretization; surprisingly, both the original system and the discretization have the same
constants of motion.

\subsection{The quadratic vector fields} Recall that $z_3$, which is the Hamiltonian of the quadratic vector fields
of the $\Upsilon$-systems, is given~by
\begin{equation*}
  z_3=\Y1n3=\sum_{\tiny\begin{array}{c}1\leqs i<j<k\leqs n\\ i,k\hbox{ odd;}\ j\hbox{ even}\end{array}}x_ix_jx_k\;.
\end{equation*}%
When $n$ is odd, $n=2m-1$, the quadratic vector field $\X_{z_3}$ is explicitly given by
\begin{equation}\label{eq:quad_vfs}
  \begin{array}{rcl}
    \dot x_{2\ell-1}&=&\displaystyle x_{2\ell-1}\left(-\sum_{j=1}^{\ell-1}x_{2j-1}+\sum_{j=\ell+1}^mx_{2j-1}\right)\quad (\ell=1,\dots,m)\;,\\
    \dot x_{2\ell}&=&\displaystyle x_{2\ell}\left(\sum_{j=1}^{\ell}x_{2j-1}-\sum_{j=\ell+1}^mx_{2j-1}\right)\quad (\ell=1,\dots,m-1)\;;
  \end{array}
\end{equation}
when $n$ is even, $n=2m$, it is given by (\ref{eq:quad_vfs}) plus one extra equation, to wit,
\begin{equation}\label{eq:quad_vfs_last}
  \dot x_{2m}=-\sum_{\tiny\begin{array}{c}1\leqs i<j< 2m\\ i\hbox{ odd;}\ j\hbox{ even}\end{array}}x_ix_j\;.
\end{equation}%
The proof of these formulas is by direct computation. Let us check the first formula in (\ref{eq:quad_vfs}):
\begin{eqnarray*}
  \pb{x_{2\ell-1},z_3}_n&=&
      \sum_{\tiny\begin{array}{c}1\leqs i<j<k\leqs n\\ i,k\hbox{ odd;}\ j\hbox{ even}\end{array}}x_i\pb{x_{2\ell-1},x_j}_nx_k\\
  &=& \sum_{\tiny\begin{array}{c}1\leqs i<2\ell<k\leqs n\\ i,k\hbox{ odd}\end{array}}x_ix_k-
      \sum_{\tiny\begin{array}{c}1\leqs i<2(\ell-1)<k\leqs n\\ i,k\hbox{ odd}\end{array}}x_ix_k\\
  &=& \sum_{\tiny\begin{array}{c}1\leqs 2\ell<k\leqs n\\ k\hbox{ odd}\end{array}}x_{2\ell-1} x_k-
      \sum_{\tiny\begin{array}{c}1\leqs i<2(\ell-1)\\ i\hbox{ odd}\end{array}}x_ix_{2\ell-1}\;.
\end{eqnarray*}

When considering the integration of the vector field $\X_{z_3}$ in dimension $n=2m$, one may first integrate the
vector field in dimension $n-1$ since (\ref{eq:quad_vfs}) is independent of $x_{2m}$; then $x_{2m}$ can be obtained
from it by simply integrating the right hand side in (\ref{eq:quad_vfs_last}), because it is also independent of
$x_{2m}$. Geometrically speaking, the vector field $\X_{z_3}$ in dimension $n=2m-1$ is a (Poisson) reduction of the
vector field $\X_{z_3}$ in dimension $n=2m$. We therefore concentrate in the sequel on the case $n=2m-1$, that is,
on the equations~(\ref{eq:quad_vfs}).

\subsection{A superintegrable subsystem}\label{par:sub_super}
A closer look at the first set of equations in (\ref{eq:quad_vfs}) reveals that they involve only the variables
$x_k$, with $k$ odd. This means that the vector field $\X_{z_3}$ projects to a vector field on $\bbR^m$, under the
map
\begin{equation}
  \begin{array}{lcccl}
    \phi&:&\bbR^{2m-1}&\to& \bbR^m\\
        & &(x_1,x_2,\dots,x_{2m-1}) &\mapsto&(x_1,x_3,\dots,x_{2m-1})\;.
  \end{array}
\end{equation}
Denoting the coordinates on $\bbR^m$ by
\begin{equation} \label{yx}
y_i=x_{2i-1},\quad \quad i=1,\ldots,m,
\end{equation}
the projected vector field is given by
\begin{equation}\label{eq:subsystem}
    \dot y_i= y_{i}\left(-\sum_{j=1}^{i-1}y_{j}+\sum_{j=i+1}^my_{j}\right)\qquad (i=1,\dots,m)\;.
\end{equation}
Such a vector field goes under the name of \emph{Lotka-Volterra system} \cite{Vol, Bog}. In fact, it is a special
Lotka-Volterra of the form $\dot y_j=\sum_i c_{ij}y_iy_j$, where the constants $c_{ij}$
satisfy $c_{ij}=-c_{ji}$. This implies we have a Poisson structure and a Hamiltonian \cite{Dam}. Notice that
the quadratic vector field (\ref{eq:quad_vfs}) of the $\Upsilon$-systems is \emph{not} of this form because the
constants $c_{ij}$ do not satisfy the skew-symmetry property. Thus the subsystem (\ref{eq:subsystem}) is a
Hamiltonian vector field, but \emph{not} with respect to the Poisson structure which
is induced from the Poisson structure on $\bbR^{2m-1}$ via the map $\phi$ (the induced Poisson structure is
trivial). Instead, consider the quadratic Poisson structure on $\bbR^m$, defined by
\begin{equation} \label{PB2}
\pb{y_i,y_j}^q:=y_iy_j,
\end{equation}
for any $1\leqs i<j\leqs m$. This bracket is distinguished notationally from the bracket (\ref{PB1}) by the
superscript $q$ (for quadratic) and we have omitted the dependence on the dimension. It is a well-known fact that this indeed defines a Poisson structure, see e.g.\ \cite[Example 8.14]{LPV}. Then it is clear that
\begin{equation} \label{eq:H}
H:=z_1=\sum_{i=1}^my_i
\end{equation}
is a Hamiltonian for the vector field (\ref{eq:subsystem}). Infinitely many copies of the system (\ref{eq:subsystem}) did also arise in the work of Bogoyavlenskij \cite[Equation 2.8]{Bog}, who provided exact solutions, and, in the $m=4$ dimensional case, gave 3 integrals. In what follows we
show, for any $m$, that this Hamiltonian system is superintegrable,
and that it is Liouville integrable.
\begin{prop}\label{prp:first_integrals}
  The Hamiltonian system (\ref{eq:subsystem}) on $\bbR^m$ admits for $1\leqs k\leqs\left[\frac{m+1}2\right]$ the
  following rational functions as constants of motion (first integrals):
  \begin{equation}\label{E:integrals}
    F_k:= \left\{ \begin{array}{ll}
    \left(y_1+y_2+\cdots+y_{2k-1}\right)\ds\frac{y_{2k+1}y_{2k+3}\ldots y_{m}}{y_{2k}y_{2k+2}\ldots y_{m-1}}&\mbox{ if}\  m\mbox{ is odd},\\
    \\
    \left(y_1+y_2+\cdots+y_{2k}\right)\ds\frac{y_{2k+2}y_{2k+4}\ldots y_m}{y_{2k+1} y_{2k+3}\ldots y_{m-1}}&\mbox{ if} \  m\mbox{ is even}.
    \end{array} \right.
  \end{equation}
\end{prop}
\begin{proof}
We assume $m$ is odd. Then $F_k$ can be written as
\begin{equation}\label{eq:F_abs}
  F_k=\sum_{\ell=1}^{2k-1}y_\ell\prod_{s=k}^{(m-1)/2}\frac{y_{2s+1}}{y_{2s}}\;.
\end{equation}%
It is easily computed from (\ref{eq:subsystem}) that
\begin{equation*}
 \sum_{\ell=1}^{2k-1}\dot y_\ell=\left(\sum_{\ell=1}^{2k-1}y_\ell\right)\left(\sum_{\ell=2k}^my_\ell\right) \hbox{ and }
  \frac{\text{d}}{\text{d}t}\left(\frac{y_{2s+1}}{y_{2s}}\right)=-\frac{y_{2s+1}}{y_{2s}}(y_{2s}+y_{2s+1})\;,
 \end{equation*}%
and so
\begin{equation*}
  \frac{\text{d}}{\text{d}t}\left(\prod_{s=k}^{(m-1)/2}\frac{y_{2s+1}}{y_{2s}}\right)=
  -\left(\prod_{s=k}^{(m-1)/2}\frac{y_{2s+1}}{y_{2s}}\right)\left(\sum_{\ell=2k}^my_\ell\right)\;.
\end{equation*}%
The fact that $\dot F_k=0$ follows at once from these formulas. The case where $n$ is even can be proven similarly.
\end{proof}
Notice that $H=F_{[(m+1)/2]}$. More constants of motion can be produced by the following trick. On $\bbR^m$ we
consider the involution $\imath$, defined by
\begin{equation*}
  \imath(y_1,y_2,\dots,y_m):=(y_m,y_{m-1},\dots,y_1)\;.
\end{equation*}%
The map $\imath$ is anti-Poisson map, since for any $i<j$,
$$
  \pb{\imath^*y_i,\imath^*y_j}^q=\pb{y_{m+1-i},y_{m+1-j}}^q=-y_{m+1-i}y_{m+1-j}=-\imath^*\pb{y_i,y_j}^q\;.
$$
Also, $H$ is invariant under this involution, $\imath^*H=H$. As a consequence,
\begin{equation*}
  \left(\imath^* F_k\right)^\cdot=\pb{\imath^* F_k,H}^q=\pb{\imath^* F_k,\imath^*H}^q=-\imath^*\pb{F_k,H}^q=0\;,
\end{equation*}%
which proves that the rational functions $G_k:=\imath^*F_k$ ($k=1,\dots,\left[\frac{m+1}2\right]$) are constants of
motion of~(\ref{eq:subsystem}). Notice that we have constructed  precisely $m-1$ different constants of motion:
when $m$ is even, all $F_k$ and $G_k$ are different, except for $F_{m/2}=H=G_{m/2}$; when $m$ is odd,
all $F_k$ and $G_k$ are different, except for $F_{(m+1)/2}=H=G_{(m+1)/2}$ and $F_1=G_1$. We note that for the 4-dimensional Lotka-Volterra system (\ref{eq:subsystem}) the integrals coincide with the ones given by
Bogoyavlenskij \cite[(7.9)]{Bog}.

\begin{prop}\label{prp:int_both}
  Let $r:=\left\lbrack\frac{m+1}2\right\rbrack$, so that $m=2r$ when $m$ is even and $m=2r-1$ when $m$ is odd. For
  $i$ and $j$, satisfying $1\leqs i,j<r$, we have
  \begin{align}
    \pb{F_i,F_j}^q&=\pb{F_i,H}^q=0\;,\label{eq:poi_fi}\\
    \pb{G_i,G_j}^q&=\pb{G_i,H}^q=0\;.\label{eq:poi_gi}
  \end{align}
Moreover, the following $m-1$ functions are independent:
  \begin{align}
    F_1,\dots,F_{r-1},G_1,\dots,G_{r-1},G_r=F_r=H,\mbox{ when $m$ is even,} \label{FGe}\\
    F_1=G_1,F_2,\dots,F_{r-1},G_2,\dots,G_{r}=F_r=H,\mbox{ when $m$ is odd.} \label{FGo}
  \end{align}
  As a consequence,
  \begin{enumerate}
    \item[(1)] The vector field (\ref{eq:subsystem}) is superintegrable;
    \item[(2)] $(\bbR^m,\PB^q,(F_1,\dots,F_{r-1},H))$ is a Liouville integrable system;
    \item[(3)] $(\bbR^m,\PB^q,(G_1,\dots,G_{r-1},H))$ is a Liouville integrable system.
  \end{enumerate}
\end{prop}
\begin{proof}
In Appendix \ref{AA} we prove that both sets of functions (\ref{FGe}) and (\ref{FGo}) are functionally
independent. Here we prove that $\pb{F_i,F_j}^q=0$ for all $i$ and $j$ satisfying $1\leqs i,j<r$. We do this by
induction on $m$.  For $m=2,3,4$ there is nothing to prove.  Assume therefore that the formula is correct for some
$m\geqs 4$; we show that it holds for $m+2$. We denote the functions $F_k$ which are constructed in dimension $m$
by $F_k^{(m)}$ and we set, as before, $r:=[(m+1)/2]$. For $k=1,\dots,r$, we have
\begin{equation}\label{eq:f_rec}
  F_k^{(m+2)}=F_k^{(m)}\frac{y_{m+2}}{y_{m+1}}\;.
\end{equation}%
Let $1\leqs i,j< r+1$. In view of the above formulas and the induction hypothesis,
\begin{eqnarray*}
  \pb{F_i^{(m+2)},F_j^{(m+2)}}^q
    &=&\pb{F_i^{(m)}\frac{y_{m+2}}{y_{m+1}},F_j^{(m)}\frac{y_{m+2}}{y_{m+1}}}^q\\
    &=&\frac{y_{m+2}}{y_{m+1}}\left(F_i^{(m)}\pb{\frac{y_{m+2}}{y_{m+1}},F_j^{(m)}}^q-F_j^{(m)}\pb{\frac{y_{m+2}}{y_{m+1}},F_i^{(m)}}^q\right)\;.
\end{eqnarray*}
The latter two brackets are zero for the following general reason: if $F$ is a function which depends only on
$y_1,\dots,y_m$, then $\pb{y_{m+2}/y_{m+1},F}=0$. To show this fact, let $1\leqs i\leqs m$. Then
\begin{equation*}
  \pb{\frac{y_{m+2}}{y_{m+1}},y_i}^q=-\frac{y_{m+2}y_i}{y_{m+1}}+\frac{y_{m+2}y_{m+1}y_i}{y^2_{m+1}}=0\;.
\end{equation*}%
This shows that $\pb{F_i^{(m+2)},F_j^{(m+2)}}^q=0$ for $1\leqs i,j< r+1$, which proves (\ref{eq:poi_fi}), since we
know from Proposition \ref{prp:first_integrals} that the functions $F_i$ are first integrals of $\X_H$, hence are
in involution with $H$. Also, since $\imath$ is an anti-Poisson map, (\ref{eq:poi_gi}) follows immediately from
(\ref{eq:poi_fi}).

This shows that in both cases ($m$ even or odd), the Hamiltonian vector field $\X_H$, which is explicitly given by
(\ref{eq:subsystem}), has $m-1$ independent constants of motion, hence is superintegrable. When $m$ is even, the
rank of the Poisson structure $\PB^q$ is equal to $m$, so that the $r$ functions $F_1,\dots,F_{r-1},H$, which are
independent and in involution, define a Liouville integrable system on $(\bbR^m,\PB^q)$. When $m$ is odd, $F_1=G_1$
is a Casimir function of $\PB^q$ and one needs for Liouville integrability, besides the Casimir function
$(m-1)/2=r-1$ independent functions in involution, so again the functions $F_1,\dots,F_{r-1},H$ define a Liouville
integrable system on $(\bbR^m,\PB^q)$. This shows property (2). Property (3) is an immediate consequence of it upon
using the involution $\imath$; in particular, the Liouville integrable systems in (2) and (3) are isomorphic.
\end{proof}
It is a classical result, due to Poisson, that the Poisson bracket of two constants of motion is again a constant
of motion. We give in the following proposition explicit formulas for the Poisson bracket of the constants of
motion $F_i$ and $G_j$ of $\X_H$, as proven in Appendix \ref{AB}.
\begin{prop}\label{prp:br_Fi_Gj}
 Let $r:=\left\lbrack\frac{m+1}2\right\rbrack$, as in Proposition \ref{prp:int_both}. For  $i$ and $j$, satisfying
 $1\leqs i,j<r$, set $\kappa:=i+j-r-1$. Then
$$
  \pb{F_i,G_j}^q=\left\{
  \begin{array}{ll}
    -F_iG_j&\kappa<0,\ m \mbox{ even,}\\
    0&\kappa<0,\ m \mbox{ odd,}\\
    -F_{r-j}G_{r-i}&\kappa\geqs0,\ m \mbox{ even,}\\
    \ds(-1)^\kappa\frac{\prod_{\ell=0}^\kappa(F_1G_r-F_{i-\kappa+\ell}G_{j-\ell})}
            {\prod_{\ell=1}^\kappa(F_1G_r+F_{i-\kappa+\ell-1}G_{j-\ell})}\quad&\kappa\geqs0,\ m \mbox{ odd.}
  \end{array}
  \right.
$$
\end{prop}

Since the vector field (\ref{eq:subsystem}) is superintegrable, it can be quite explicitly integrated. This can be
done as follows. For $j=1,\dots,m$, define $u_j:=y_1+y_2+\cdots+y_j$. Then $H=u_m$. It is easy to see that in terms
of the functions $u_i$, the Poisson structure is given by
\begin{equation*}
  \pb{u_i,u_j}^q=u_i(u_j-u_i)\;,\qquad i<j\;,
\end{equation*}%
in particular the Hamiltonian vector field (\ref{eq:subsystem}) is in terms of these coordinates given by
\begin{equation}\label{eq:final_ode}
  \dot u_i=\pb{u_i,H}^q=u_i(H-u_i) \;.
\end{equation}%
Thus, the variables $u_i$ provide a separation of variables and each one of the variables satisfies the same
differential equation (\ref{eq:final_ode}); its explicit integration is immediate, and can be found in \cite{Bog}.

\subsection{Superintegrability and non-commutative integrability of the quadratic vector fields}\label{par:super}
We now show how the superintegrability of the $m$-dimensional Lotka-Volterra system, studied in Section \ref{par:super}, extend to the superintegrability (resp.\ non-commutative integrability) of the quadratic vector field of the $\Upsilon$-system in
dimension $n=2m-1$ (resp.\ $n=2m$).
\begin{prop}\label{prp:quad_odd_super}
  When $n$ is odd, the quadratic vector field of the $n$-dimensional $\Upsilon$-system is superintegrable.
\end{prop}
\begin{proof}
Write $n=2m-1$. Suppose first that $m$ is odd, set $m=2r-1$  and consider the following $n-1$ functions:
$$
  z_1,z_3,\ldots,z_n,F_1,F_2,\ldots,F_{r-1},G_2,G_3,\ldots,G_{r-1}\;.
$$
Since the quadratic vector field is the Hamiltonian vector field $\X_{z_3}$, Theorem \ref{thm:Y_invol} implies that
the functions $z_1,z_3,z_5,\ldots,z_n$ are in involution with $z_3$, hence are constants of motion
of~$\X_{z_3}$. Furthermore, according to Proposition \ref{prp:int_both}, the above functions $F_i$ and $G_j$, with (\ref{yx}), are
constants of motion of $\Pb H^q$, which is the projected vector field of $\X_{z_3}$, hence they are, viewed as
functions on $\bbR^n$, constants of motion of $\X_{z_3}$. It can be shown that these $n-1$ functions are
functionally independent (a proof is given in Appendix \ref{AA}). This shows that the vector field $\X_{z_3}$ is
superintegrable when $m$ is odd. The proof in case $m$ is even, $m=2r$, is the same; in this case one uses the
following $n-1$ functions:
$$
  z_1,z_3,\ldots,z_n,F_1,F_2,\ldots,F_{r-1},G_1,G_2,\ldots,G_{r-1}\;.
$$
\end{proof}
\begin{prop}\label{prp:quad_even_super}
  When $n$ is even, the quadratic vector field of the $n$-dimensional $\Upsilon$-system is a non-commutative
  integrable system of rank $2$ on $(\bbR^n,\PB_n)$.
\end{prop}
\begin{proof}
Write $n=2m$ and suppose first that $m$ is odd, $m=2r-1$. Consider the following  $n-2$ functions:
$$
  z_1,z_3,\ldots,z_{n-1},F_1,F_2,\ldots,F_{r-1},G_2,G_3,\ldots,G_{r-1}\;.
$$
As in the proof of Proposition \ref{prp:quad_odd_super}, $z_3$ is in involution with each one of these
functions. Since $z_1=G_r=F_r=H$, we have that $z_1$ is also in involution with all these functions. Since these
$n-2$ functions are independent, as is shown in Appendix \ref{AA}, and since the Hamiltonian vector fields of $z_1$
and $z_3$ are independent (at a generic point of $\bbR^n$), this shows that these $n-2$ functions define a
non-commutative integrable system of rank two.
The proof in case $m$ is even, $m=2r$, is the same; one uses in this case the following $n-2$ functions:
$$
  z_1,z_3,\ldots,z_{n-1},F_1,F_2,\ldots,F_{r-1},G_1,G_2,\ldots,G_{r-1}\;.
$$
\end{proof}

\subsection{Kahan discretization}
Kahan discretization was introduced as an unconventional discretisation method in \cite{Kah,KL}.
It seems to have quite remarkable properties in the sense of preserving geometric structures \cite{PPS,CMOQ}.
In this section we consider the Kahan discretization of the quadratic Hamiltonian system (\ref{eq:subsystem}).
The Kahan discretization of (\ref{eq:subsystem}) with step size $2\e$ is given by
\begin{align*}
  \yt_1-y_1&=\e y_1\left(\yt_2+\yt_3+\ldots +\yt_m\right)+\e\yt_1\left(y_2+y_3+\ldots+y_{m}\right)\;,\\
  \yt_2-y_2&=\e y_2\left(-\yt_1+\yt_3+\ldots+\yt_m\right)+\e \yt_2\left(-y_1+y_3+\ldots+y_m\right)\;,\\
  \vdots &\qquad\qquad\qquad\qquad\qquad \vdots \\
  \yt_m-y_m&=\e y_m\left(-\yt_1-\yt_2-\ldots -\yt_{m-1}\right)+\e\yt_{m}\left(-y_1-y_2-\ldots-y_{m-1}\right)\;.
\end{align*}
In what follows we introduce the variables
\begin{equation}\label{eq:u}
u_i:=y_1+y_2+\cdots+y_i,\qquad \tilde u_i:=\yt_1+\yt_2+\cdots+\yt_i.
\end{equation}
Thus, $u_m=H$, cf. (\ref{eq:H}), and it follows at once from summing up the above equations that
$\tilde u_m=u_m$, i.e., that $H$ is an integral of the discrete map
\begin{equation} \label{E:map}
  (y_1,y_2,\ldots,y_m)\mapsto (\yt_1,\yt_2,\ldots,\yt_m) .
\end{equation}
We will prove that, more generally, all the integrals given by (\ref{E:integrals}) are also integrals for this
discrete system. To do this, we use the following lemma.
\begin{lemma}\label{lma:discrete}
The Kahan discretisation of (\ref{eq:subsystem}) is explicitly given by, for $1\leq j\leq m$,
  \begin{equation}\label{E:map_explicit}
  \yt_j=y_j\frac{(1-\e H)(1+\e H)} {\left(1-\e H+2\e u_{j-1}\right)\left(1-\e H+2\e u_j\right)}\;,
  \end{equation}
\end{lemma}
where $u_j$, and $H$, are given by (\ref{eq:u}), and (\ref{eq:H}), respectively.
\begin{proof}
Summing up the first $j$ equations of the above Kahan discretization, we get, upon using $H=u_m=\tilde u_m$,
\begin{equation*}
  \tilde u_j-u_j=\e u_j(H-\tilde u_j)+\e\tilde u_j(H-u_j)\;,
\end{equation*}
which can be solved linearly as
\begin{equation}\label{eq:ut_sol}
  \tilde u_j=u_j\frac{1+\e H}{1-\e H+2\e u_j}\;.
\end{equation}%
The formula (\ref{E:map_explicit}) follows at once from it, since $y_j=u_j- u_{j-1}$ and $\tilde y_j=\tilde
u_j-\tilde u_{j-1}$.
\end{proof}
We remark that equation (\ref{eq:ut_sol}) is the Kahan discretization of equation (\ref{eq:final_ode}).
\begin{prop}\label{prp:kahan_integrals}
  The rational functions $F_k$, defined in proposition \ref{prp:first_integrals}, and the functions
  $G_k:=\imath^*F_k$ are first integrals of the Kahan discretization \ref{E:map_explicit}.
\end{prop}
\begin{proof}
We give the proof for $m$ odd; it is essentially the same for $m$ even. First, observe from Lemma
\ref{lma:discrete} that
\begin{equation}\label{eq:yquot}
  \frac{\yt_{2s+1}}{\yt_{2s}}=\frac{y_{2s+1}}{y_{2s}}\;\frac{1-\e H+2\e u_{2s-1}}{1-\e H+2\e u_{2s+1}}\;.
\end{equation}%
Therefore, using the formula (\ref{eq:F_abs}) for $F_k$, (\ref{eq:ut_sol}) and (\ref{eq:yquot}) we get that
\begin{eqnarray*}
  \tilde F_k&=& \tilde u_{2k-1}\prod_{s=k}^{(m-1)/2}\frac{\yt_{2s+1}}{\yt_{2s}}\\
            &=&u_{2k-1}\frac{1+\e H}{1-\e H+2\e u_{2k-1}}\prod_{s=k}^{(m-1)/2}
                \left(\frac{y_{2s+1}}{y_{2s}}\;\frac{1-\e H+2\e u_{2s-1}}{1-\e H+2\e u_{2s+1}}\right)\\
            &=&u_{2k-1}\frac{1+\e H}{1-\e H+2u_m}\prod_{s=k}^{(m-1)/2}\frac{y_{2s+1}}{y_{2s}}\\
            &=&u_{2k-1}\prod_{s=k}^{(m-1)/2}\frac{y_{2s+1}}{y_{2s}}=F_k\;,
\end{eqnarray*}
where the last line was obtained by using that $H=u_m$. This shows that each $F_k$ is a first integral. Since the
discrete system is invariant under the map $\imath$ (upon replacing $\e$ by $-\e$), each $G_k$ is also a first
integral.
\end{proof}
\begin{prop}\label{prp:kahan_poisson}
  The Kahan discretization of (\ref{eq:subsystem}) is a Poisson map.
\end{prop}
\begin{proof}
Notice that in terms of the coordinates $u_i$, the Poisson bracket is given by $\pb{u_i,u_j}^q=u_i(u_j-u_i)$, for
$i<j$; in particular, $\pb{u_i,H}^q=u_i(H-u_i)$.  It therefore suffices to show that $\pb{\tilde u_i,\tilde
  u_j}^q=\tilde u_i(\tilde u_j-\tilde u_i)$ for $i<j$. Notice that $\tilde u_i$ depends only on $u_i$ and $H$, since
\begin{equation}
  \tilde u_i=u_i\frac{1+\e H}{1-\e H+2\e u_i}\;.
\end{equation}%
We have
\begin{equation*}
  \pp{\tilde u_i}{u_i}=\frac{1-\e^2 H^2}{(1-\e H+2\e u_i)^2}\;,\qquad
  \pp{\tilde u_i}{H}=\frac{2\e u_i(1+\e u_i)}{(1-\e H+2\e u_i)^2}\;.
\end{equation*}%
It follows that
\begin{eqnarray*}
  \pb{\tilde u_i,\tilde u_j}^q
  &=& \pp{\tilde u_i}{u_i}\pp{\tilde u_j}{u_j}\pb{u_i,u_j}^q+\pp{\tilde u_i}{u_i}\pp{\tilde u_j}{H}\pb{u_i,H}^q+
       \pp{\tilde u_i}{H}\pp{\tilde u_j}{u_j}\pb{H,u_j}\\
  &=& \frac{(1-\e^2H^2)u_i(u_j-u_i)}{(1-\e H+2\e u_i)^2(1-\e H+2\e u_j)^2}(1-\e^2H^2+2\e(1+\e H)u_j)\\
  &=& \frac{(1+\e H)^2(1-\e H)u_i(u_j-u_i)}{(1-\e H+2\e u_i)^2(1-\e H+2\e u_j)}\\
  &=& \tilde u_i(\tilde u_j-\tilde u_i)\;.
\end{eqnarray*}
as was to be shown.
\end{proof}
For a discrete map in dimension $n$ the existence of $n-1$ integrals is not enough to claim (super)integrability.
\begin{defn}
An $n$-dimensional map is {\it superintegrable} if it has $n-1$ constants of motion and it is measure preserving.
A $n=2r+s$-dimensional map on a Poisson manifold, which respects the Poisson structure of rank $2r$ is {\em Liouville integrable}
if there are $n-r=r+s$ functionally independent constants of motion in involution, cf. \cite{BRSG,Mea,Ves}.
\end{defn}

It is known that for a symplectic map $L:\rm{y}\mapsto \rm{\tilde{y}}$ with structure matrix $\Omega$ we have
$$
dL. \Omega. dL^{T}=\widetilde{\Omega},
$$
where $dL$ is the Jacobian matrix of $L$. It yields $\left(\rm{det}(dL)\right)^2=\rm{det}(\widetilde{\Omega})/\rm{det}(\Omega)$.
By calculating the determinant of our structure matrix when $m$ is even, we obtain the following result.
The result also holds for odd $m$, which is why we provide a direct proof, i.e. without assuming symplecticity.

\begin{prop}\label{prp:kahan_measure}
  The Kahan discretization (\ref{E:map_explicit}) is measure preserving, with measure
	$$
	\frac{1}{y_1y_2\ldots y_m}dy_1 \wedge dy_2 \wedge \cdots \wedge dy_m.
	$$
\end{prop}
\begin{proof}
It is easy to see that 
\begin{equation}
\label{E:Jaco}
\frac{\partial(\tilde{y}_1,\tilde{y}_2,\ldots,\tilde{y}_m)}{\partial (y_1,y_2,\ldots,y_m)}=\frac{\partial(\tilde{y}_1,\tilde{y}_2,\ldots,\tilde{y}_m)}
{\partial (u_1,u_2,\ldots,u_m)}.\frac{\partial(u_1,u_2,\ldots,u_m)}{\partial (y_1,y_2,\ldots,y_m)}. 
\end{equation}
Since $\tilde y_i=\tilde u_{i}-\tilde{u}_{i-1}$, we have
$$
\frac{\partial(\tilde{y}_1,\tilde{y}_2,\ldots,\tilde{y}_m)}
{\partial (u_1,u_2,\ldots,u_m)}
=
\frac{\partial(\tilde{u}_1,\tilde{u}_2,\ldots,\tilde{u}_m)}
{\partial (u_1,u_2,\ldots,u_m)}
-\frac{\partial(0,\tilde{u}_1,\tilde{u}_2,\ldots,\tilde{u}_{m-1})}
{\partial (u_1,u_2,\ldots,u_m)}:=A
$$
Entries of $A$ are obtained by direct calculation and given as follow
\begin{equation}
A[i,j]=\left\{ \begin{array}{ll}
\frac{\partial\tilde u_i}{\partial u_i}=\frac{1-\epsilon^2 u_m^2}{(1-\epsilon u_m+2 \epsilon u_i)^2}&\ \mbox{if}\ j=i,\\
-\frac{\partial\tilde u_{i-1}}{\partial u_{i-1}}=-\frac{1-\epsilon^2 u_m^2}{(1-\epsilon u_m+2 \epsilon u_{i-1})^2}&\ \mbox{if}\ j=i-1,\\
\frac{\partial\tilde u_i}{\partial u_m}-\frac{\partial\tilde u_{i-1}}{\partial u_m}
=\frac{2\epsilon u_i(1+\epsilon u_i)}{(1-\epsilon u_m+2\epsilon u_i)^2}-\frac{2\epsilon u_{i-1}(1+\epsilon u_{i-1})}{(1-\epsilon u_m+2\epsilon u_{i-1})^2}
&\ \mbox{if}\ j=m,\\
0&\ \mbox{otherwise}.\\
\end{array}
\right.
\end{equation}
To calculate the determinant of A, we divide the $j^{\rm th}$ column by $\frac{1-\epsilon^2 u_m^2}{(1-\epsilon u_m+2 \epsilon u_i)}$ for all $j<m$ 
 and then adding the first $i_1$ rows to the $i^{\rm th}$ row. We obtain  an upper triangular matrix with $1$ on the diagonal.
 Therefore, one gets
 \begin{align}
 {\rm det} (A)&=\frac{(1-\epsilon^2 u_m^2)^{m-1}}{(1-\epsilon u_m+2 \epsilon u_1)^2 (1-\epsilon u_m+2 \epsilon u_2)^2\ldots (1-\epsilon u_m+2 \epsilon u_{m-1})^2}
 \label{E:Jaco1}\\
 &=\frac{\tilde{y}_1 \tilde{y}_2\ldots \tilde{y}_{m}}{y_1y_2\ldots y_m}\label{E:Jaco2}
 \end{align}
On the other hand,
the determinant of $\frac{\partial(u_1,u_2,\ldots,u_m)}{\partial (y_1,y_2,\ldots,y_m)}$ in~\eqref{E:Jaco} is $1$. Thus,
$$
\frac{d\tilde{y}_1\we \cdots\we d\tilde{y}_m}{\tilde{y}_1 \cdots \tilde{y}_m} = \frac1{\tilde{y}_1 \cdots \tilde{y}_m}\frac{\p(\tilde{y}_1,\ldots,\tilde{y}_m)}{\p(y_1,\ldots,y_m)} dy_1\we \cdots \we dy_m = \frac{dy_1\we \cdots \we dy_m}{y_1\cdots y_m},
$$
which shows the Kahan discretization (\ref{E:map_explicit}) is measure preserving, with the given measure.
\end{proof}

As a direct consequence of Propositions \ref{prp:int_both}, \ref{prp:kahan_integrals}, \ref{prp:kahan_poisson}, and
\ref{prp:kahan_measure} we get the following result.
\begin{prop}
The Kahan discretization (\ref{E:map_explicit}) is both superintegrable, and Liouville integrable.
\end{prop}

Finally we'd like to remark that due to the preservation of both the Poisson structure, and the Hamiltonian,
the Kahan discretization (\ref{E:map_explicit}) is the time advance map
for the exact Hamiltonian system (\ref{eq:subsystem}) up to a reparametrization of time \cite{ZM}.
%

\section*{Acknowledgments}
This research was supported by the Australian Research Council and by the Centre of Excellence for Mathematics and Statistics of Complex Systems (MASCOS). All authors are grateful for the hospitality of the Isaac Newton Institute during the follow-up meeting 'Discrete Integrable Systems' (July 2013). PV would like to thank the Department of Mathematics and Statistics of La Trobe University for supporting his visit (Dec. 2013).

\appendix

\section{Functional independence of the integrals of the quadratic vector fields}
\label{AA}
We first show that when $m=2r$ is even, the $m-1$ functions $F_1,\dots,F_{r-1},$ $G_1,\dots,G_{r-1},H=F_r=G_r$ are
(functionally) independent, as was asserted in the proof of Proposition \ref{prp:int_both}. Let us write $\bf 1$ as
a shorthand for the point $(1,1,\dots,1)\in\bbR^{2r}$. At this point, we have that
\begin{equation}\label{eq:der_at_1}
  \pp{F_k}{y_i}({\bf 1})=\pp{G_k}{y_{2r+1-i}}({\bf 1})= \left\{ 
  \begin{array}{cl}
    1 &\mbox{ if }1\leqs i\leqs 2k\;,\\
    2k&\mbox{ if }2k<i\leqs 2r\mbox{ and }i\mbox{ is even},\\
    -2k&\mbox{ if }2k<i\leqs 2r\mbox{ and }i\mbox{ is odd}.
  \end{array} \right.
\end{equation}
Define rational functions $K_1,\dots K_{2r-1}$ by
\begin{equation*}
  K_j:=\left\{
  \begin{array}{cl}
    F_{(j-1)/2}-H &\mbox{ if } j \mbox{ is odd},\\
    G_{r+1-j/2}-(r+1-j/2)G_1 &\mbox{ if } j \mbox{ is even},
  \end{array} \right.
\end{equation*}
where $F_0:=0$, so that $K_1=-H$. We show that the Jacobian of these functions is of maximal rank ($2r-1$) at ${\bf
1}$. On the one hand, (\ref{eq:der_at_1}) implies that
\begin{equation*}
  \pp{K_j}{y_i}({\bf 1})=0 \mbox{ for } \left\{
  \begin{array}{ll}
    j\mbox{ odd},\ i<j\;,\\
    j\mbox{ even},\ i<j-1\;.
  \end{array} \right.
\end{equation*}%
On the other hand, for $i=1,\dots,r-1$, the matrices 
\begin{equation*}
  \renewcommand{\arraystretch}{1.5}
  \begin{pmatrix}
    \pp{K_{2i-1}}{y_{2i-1}}({\bf 1})&\pp{K_{2i-1}}{y_{2i}}({\bf 1})\\
    \pp{K_{2i}}{y_{2i-1}}({\bf 1})&\pp{K_{2i}}{y_{2i}}({\bf 1})
  \end{pmatrix}
  =
  \begin{pmatrix}
    1-2i&2i-3\\
    2i-1-2r&2r-2i+3  
  \end{pmatrix}
\end{equation*}%
are all non-singular and $\pp{K_{2r-1}}{y_{2r-1}}({\bf 1})=-(2r-1)\neq0$. It follows that the Jacobian of the
functions $K_1,\dots,K_{2r-1}$, and hence of the functions $F_1,\dots,F_{r-1},G_1,\dots,G_{r-1},H=F_r=G_r$, is of
maximal rank at the point ${\bf 1}$. This shows that the latter functions are functionally independent on
$\bbR^{2r}$. The proof that when $m=2r-1$ is odd, the $m-1$ functions $F_1=G_1,F_2,$ $\dots,F_{r-1},G_2,\dots,G_{r}=F_{r}=H$ are
functionally independent goes along the same lines. 

As was stated in Section \ref{sec:subsystem}\ref{par:super}, the above rational functions $K_i$ remain functionally
independent when they are viewed as functions on $\bbR^n$ and the odd polynomials $z_3,z_5,\dots$ are added to
them; here $n=2m-1$ or $n=2m$, depending on whether $n$ is even (Proposition \ref{prp:quad_odd_super}) or odd
(Proposition \ref{prp:quad_even_super}). Since the functions $K_i$ depend only on the variables $y_i=x_{2i-1}$,
i.e., are independent of the variables $x_{2i}$, we only need to verify that the Jacobian determinant of the
polynomials $z_3,z_5,\dots$ with respect to the variables $x_2,x_4,\dots$ is non-zero (at one point at least).
Precisely, when $n$ is odd (resp.\ $n$ is even) one needs to check that the rank of the following Jacobian matrices
is maximal:
$$
  \frac{\partial(z_3,z_5,\dots,z_n)}{\partial (x_2,x_4,\ldots, x_{n-1})}\;,\qquad\hbox{resp.\ }\qquad
  \frac{\partial(z_3,z_5,\dots,z_{n-1})}{\partial (x_2,x_4,\ldots, x_{n-2})}\;.
$$
The proof is very similar to the proof of Proposition \ref{prp:independent}, which shows that the Jacobian matrix 
$$
  \frac{\partial(z_1,z_2,\dots,z_n)}{\partial (x_1,x_2,\ldots, x_{n})}
$$
has maximal rank.

\section{The brackets between functions $F$ and $G$}
\label{AB}
We outline the proof of Proposition~\ref{prp:br_Fi_Gj}, which proceeds by induction.
\begin{proof} For $m<3$ there is nothing to be checked, while for $m=3$ it is
contained in Proposition~\ref{prp:int_both}. For $m=4$ we only need to check that $\pb{F_1,G_1}^q=-F_1G_1$, which is
easily done with the following explicit formulas: $F_1=(y_1+y_2)y_4/y_3$ and $G_1=(y_3+y_4)y_1/y_2$. Assuming that
the formulas hold for some $m\geqs 4$ one shows that they also hold for $m+2$. As in the proof of
Proposition~\ref{prp:int_both}, we denote the functions $F_k$ and $G_k$ which are constructed in dimension $m$ by
$F_k^{(m)}$ and $G^{(k)}$ and we set, as before, $r:=[(m+1)/2]$. We have, besides (\ref{eq:f_rec}),
\begin{equation*}
  G_j^{(m+2)}=G_{j-1}^{(m)}+(y_{m+1}+y_{m+2})\prod_{k=1}^{r-j+1}\frac{y_{2k-1}}{y_{2k}}\;,
\end{equation*}%
and so, in order to compute $\pb{F_i^{(m+2)},G_j^{(m+2)}}^q$, we only need to compute the following types of Poisson
brackets:
\begin{align*}
  &\pb{F_i^{(m)},G_{j-1}^{(m)}}^q,\mbox{ which is given by the induction hypothesis;}\\
  &\pb{F_i^{(m)},y_{m+1}+y_{m+2}}^q=F_i^{(m)}(y_{m+1}+y_{m+2}),\mbox{ see below;}\\
  &\pb{F_i^{(m)},\prod_{k=1}^{r-j+1}\frac{y_{2k-1}}{y_{2k}}}^q,\mbox{ see below;}\\
  &\pb{\frac{y_{m+2}}{y_{m+1}},G_{j-1}^{(m)}}^q=0,\mbox{ since $G_{j-1}^{(m)}$ is independent of $y_{m+1},y_{m+2}$;}\\
  &\pb{\frac{y_{m+2}}{y_{m+1}},y_{m+1}+y_{m+2}}^q=-\frac{y_{m+2}}{y_{m+1}}(y_{m+1}+y_{m+2});\\
  &\pb{\frac{y_{m+2}}{y_{m+1}},\prod_{k=1}^{r-j+1}\frac{y_{2k-1}}{y_{2k}}}^q=0,\mbox{ since each
    $\frac{y_{2k-1}}{y_{2k}}$ is independent of $y_{m+1},y_{m+2}$.}
\end{align*}
In order to show that $\pb{F_i^{(m)},y_{m+1}+y_{m+2}}^q=F_i^{(m)}(y_{m+1}+y_{m+2})$, it suffices to observe that
$\pb{y_i,y_{m+1}+y_{m+2}}^q= y_i(y_{m+1}+y_{m+2})$ when $i\leqs m$ and that~$F_i^{(m)}$ is homogeneous of degree 1 and
depends on the variables $y_1,\dots,y_m$ only. 
Finally, when $m$ is even,
\begin{equation*}
  \pb{F_i^{(m)},\prod_{k=1}^{r-j+1}\frac{y_{2k-1}}{y_{2k}}}^q=
  \left\{\begin{array}{ll}
    \ds-F_i^{(m)}\prod_{k=1}^{r-j+1}\frac{y_{2k-1}}{y_{2k}} &i+j\leqs r+1\;,\\
    \ds-F_{r+1-j}^{(m)}\prod_{k=1}^i\frac{y_{2k-1}}{y_{2k}} &i+j>r+1\;,
  \end{array}\right.
\end{equation*}%
as follows at once from 
\begin{equation*}
  \pb{F_i^{(m)},\frac{y_{2k-1}}{y_{2k}}}^q=
  \left\{\begin{array}{cc}
    \ds-\frac{y_{2k-1}}{y_{2k}}(y_{2k}+y_{2k-1})\prod_{s=i+1}^{r}\frac{y_{2s}}{y_{2s-1}}\quad &k\leqs i\;,\\
    \ds0&k>i\;;
  \end{array}\right.
\end{equation*}%
when $m$ is odd,
\begin{equation*}
  \pb{F_i^{(m)},\prod_{k=1}^{r-j+1}\frac{y_{2k-1}}{y_{2k}}}^q=
  \left\{\begin{array}{ll}
    0 &i+j\leqs r+1\;,\\
    \ds-\sum_{k=1}^{2i-2}\frac{F_1y_k}{y_{2i-1}} \prod_{s=1}^{i+j-r-2}\frac{y_{2i-2s}}{y_{2i-2s-1}} &i+j>r+1\;.
  \end{array}\right.
\end{equation*}%
as follows at once from 
\begin{equation*}
  \pb{F_i^{(m)},\frac{y_{2k-1}}{y_{2k}}}^q=
  \left\{\begin{array}{cc}
    \ds-\frac{y_{2k-1}}{y_{2k}}(y_{2k}+y_{2k-1})\prod_{s=i+1}^{r}\frac{y_{2s-1}}{y_{2s-2}}\quad &k< i\;,\\
    \ds\frac{y_{2i-1}}{y_{2i}}\sum_{j=1}^{2i-2}y_j\prod_{s=i+1}^{r}\frac{y_{2s-1}}{y_{2s-2}}\quad &k=i\;,\\
    \ds0&k>i\;.
  \end{array}\right.
\end{equation*}%
Each one of these formulas is proven easily by direct computation.
\end{proof}
\section{Lax pairs for the $\Upsilon$-systems}
\label{AC}
We prove in this appendix that the Hamiltonian vector fields defined in Section
\ref{sec:Y_systems}\ref{parg:integrability} can be written in the Lax form $\dot L=[L,B]$, where the matrices $L$
and $B$ were given in Section \ref{sec:Y_systems}\ref{parg:lax}. We do this for the systems associated to $\F'$;
for the case of $\F$, the proof is very similar. We start with the even case, $n=2m$. Recall that the Lax operator
$L=L'_{2m}$ is in this case given by (\ref{eq:lax_n_even_z_even}) and that the matrix $B=B'_{2m,k}$ is, for even
$k$ given by~(\ref{eq:B2m2k}). Recall also (from Section \ref{sec:Y_systems}\ref{parg:lax}) that this matrix $L$ is
invertible and that its inverse is given by the tridiagonal matrix $M$, with entries
$m_{ij}:=x_j^{-1}(\delta_{i,j+1}-\delta_{j,i+1})$. When $L$ satisfies the above Lax equation, then its inverse $M$
satisfies the Lax equation $\dot M=[M,B]$ and vice versa. It is therefore sufficient to prove the latter Lax
equation. We first compute its left hand side. Since the polynomials $\Y abr$ depend linearly on the variables
$x_k$, one easily computes from (\ref{def:upsilon}) that
\begin{equation}\label{E:derivative}
  \frac{\partial z_k}{\partial x_i}=\sum_{\substack{0\leqs r<k \\ r+1\equiv i}}\Upsilon^{1,i-1}_r\Upsilon^{i+1,n}_{k-r-1}\;.
\end{equation}
Therefore, we obtain for the Hamiltonian vector field associated to $z_{k}$,
\begin{align*}
  \dot{x_i}&=\{x_i,z_{k}\}=\frac{\partial z_{k}}{\partial x_{i+1}}-\frac{\partial z_{k}}{\partial x_{i-1}}=
    \sum_{\substack{0\leqs r<k \\ r\equiv i}} \left(\Upsilon^{1,i}_r\Upsilon^{i+2,n}_{k-r-1}-\Upsilon^{1,i-2}_r\Upsilon^{i,n}_{k-r-1}\right)\;;\\
  \dot m_{ij}&=\frac{\delta_{j,i+1}-\delta_{i,j+1}}{x_j^2} 
    \sum_{\substack{0\leqs r<k \\ r\equiv j}} \left(\Upsilon^{1,j}_r\Upsilon^{j+2,n}_{k-r-1}-\Upsilon^{1,j-2}_r\Upsilon^{j,n}_{k-r-1}\right)\;.
\end{align*}
This formula is to be compared with the $(i,j)$-th entry of the commutator $[M,B]$, i.e., we need to show that 
\begin{equation}\label{eq:Lax_gen}
  \dot m_{ij}=\frac{b_{i-1,j}}{x_{i-1}}- \frac{b_{i+1,j}}{x_{i+1}} -\frac{b_{i,j+1}}{x_{j}} + \frac{b_{i,j-1}}{x_{j}}\;.
\end{equation}%
This is obvious when $i>j+1$ and when $i=j$, because in these cases all terms in (\ref{eq:Lax_gen}) are zero. Let
us first show that the right hand side of (\ref{eq:Lax_gen}) is also zero when $i<j-1$. For simplicity, we assume
that $i\neq1$ and that $j\neq n$. If we substitute the values of the $b_{ij}$ and we collect on the one hand the
first two terms and on the other hand the last two terms of (\ref{eq:Lax_gen}), then we get
\begin{equation*}
   \sum_{\substack{2\leqs r\leqs k-2 \\ r\equiv i+1}}\left(\Y 1i{r-1}-\Y 1{i-2}{r-1}\right)\Y{j+1}n{k-2-r}+
   \frac{x_i}{x_j}\sum_{\substack{1\leqs r\leqs k-3 \\ r\equiv i}}\Y 1{i-1}{r-1}\left(\Y{j+2}n{k-2-r}-\Y{j}n{k-2-r}\right)\;.
\end{equation*}%
Using the first recursion relation in (\ref{eq:Y_rec}) on the first term  and the second
recursion relation in (\ref{eq:Y_rec}) on the second term, we find that both terms cancel out. We next consider the case $i=j+1$. For
simplicity we suppose that $j\neq 1$. We need to prove that $(b_{j,j}-b_{j+1,j+1})/x_j=\dot m_{j+1,j}$. Written
out, it means that we need to prove that
\begin{equation*}
  x_j\sum_{\substack{0\leqs r\leqs k-2 \\ r\equiv j}}\Y1{j-2}r\Y{j+1}n{k-r-2}-
  x_j\sum_{\substack{0\leqs r\leqs k-2 \\ r\equiv j+1}}\Y1{j-1}r\Y{j+2}n{k-r-2}+
  \sum_{\substack{0\leqs r<k \\   r\equiv j}}\left(\Y 1jr\Y{j+2}n{k-r-1}-\Y 1{j-2}r\Y{j}n{k-r-1}\right)
\end{equation*}%
is zero. If we combine the second and third term, and we use the second
recursion relation in (\ref{eq:Y_rec}) to simplify the sum of the first and last term, we get
\begin{equation}\label{eq:i=j+1}
  \sum_{\substack{1\leqs r\leqs k-1 \\ r\equiv j}}\left(\Y1{j}r-x_j\Y1{j-1}{r-1}\right)\Y{j+2}n{k-r-1}-
  \sum_{\substack{1\leqs r\leqs k-1 \\ r\equiv j}}\Y1{j-2}r\Y{j+2}n{k-r-1}\;.  
\end{equation}%
In fact, when $j$ is even, one gets two other terms, to wit $(\Y1j0-\Y1{j-2}0)\Y{j+2}n{k-1}$, but they cancel out
(recall that we supposed that $j\neq 1$). The fact that (\ref{eq:i=j+1}) equals zero follows at once from the first
recursion relation in (\ref{eq:Y_rec}). To finish, we consider the remaining case $i=j-1$. We need to prove that
$b_{i-1,i+1}/x_{i-1}- b_{i+1,i+1}/x_{i+1} -(b_{i,i+2}- b_{i,i})/x_{i+1}=\dot m_{i,i+1}$. Written out, it means
      that we need to show that the following sum of six terms is zero:
\begin{eqnarray*}
  &-&x_{i+1}^2\sum_{\substack{1\leqs r\leqs k-2 \\ r\equiv i+1}}\Y1{i-2}{r-1}\Y{i+2}n{k-2-r}-
  -x_{i+1}\sum_{\substack{1\leqs r\leqs k-2 \\ r\equiv i+1}}\Y1{i-1}{r}\Y{i+2}n{k-2-r}+
  x_ix_{i+1}\sum_{\substack{1\leqs r\leqs k-2 \\ r\equiv i}}\Y1{i-1}{r-1}\Y{i+3}n{k-2-r}\\
  &+& x_{i+1}\sum_{\substack{1\leqs r\leqs k-2 \\ r\equiv i}}\Y1{i-2}{r}\Y{i+1}n{k-2-r}-
  \sum_{\substack{1\leqs r< k \\ r\equiv i+1}}\Y1{i+1}{r}\Y{i+3}n{k-1-r}+
  \sum_{\substack{1\leqs r<k \\ r\equiv i+1}}\Y1{i-1}{r}\Y{i+1}n{k-1-r}\;.
\end{eqnarray*}
As before, we combine the terms in pairs and apply the second recursion relation in (\ref{eq:Y_rec}): terms 1 and
4 yield term I below, terms 2 and 6 yield II and terms 3 and 5 yield III (to obtain III one uses the recursion
relation twice)\footnote{For integers $a,b$ we use the Kronecker-like notation $\delta_{a\equiv b}$, which is 1
  when $a\equiv b$ (modulo 2) and 0 otherwise.}:
\begin{eqnarray*}
  \textrm{I}&=&x_{i+1}\sum_{\substack{1\leqs r\leqs k-1 \\ r\equiv i+1}}\Y1{i-2}{r-1}\Y{i+3}n{k-1-r}\;,\qquad
  \textrm{II}= \sum_{\substack{1\leqs r\leqs k-2 \\ r\equiv i+1}}\Y1{i-1}{r}\Y{i+3}n{k-1-r}+\Y1{i-1}{k-1}\delta_{i\equiv0}\\
  \textrm{III}&=&-\sum_{\substack{2\leqs r\leqs k-1 \\ r\equiv i+1}}\Y1{i}{r}\Y{i+3}n{k-1-r} 
    -x_{i+1}\sum_{\substack{2\leqs r\leqs k-1 \\ r\equiv i+1}}\Y1{i-2}{r-1}\Y{i+3}n{k-1-r}
    -\Y1{i+1}0\Y{i+3}n{k-1}\delta_{i\equiv1}    -\Y1{i+1}1\Y{i+3}n{k-2}\delta_{i\equiv0}\;.
  \end{eqnarray*}
If we add up $\textrm{II}$ with the first and third term in $\textrm{III}$, only the single boundary term
$\Y1{i-1}1\Y{i+3}n{k-2}\delta_{i\equiv0}$ remains. Similarly, adding up $\textrm{I}$ and the second term in
$\textrm{III}$ leads to the single boundary term $x_{i+1}\Y1{i-2}0\Y{i+3}n{k-2}\delta_{i\equiv0}$. So it remains to
be shown that the sum of these terms with the last term of III is zero, i.e., if $i$ is even then
\begin{equation*}
  \left(\Y1{i-1}1+x_{i+1}\Y1{i-2}0-\Y1{i+1}1\right)\Y{i+3}n{k-2}=0\;.
\end{equation*}%
That this is so follows at once from  $\Y1{i-2}0=1$ and $\Y1{i+1}1=z_1^{(i+1)}=x_1+x_3+\cdots+x_{i+1}$.
This proves the Lax equations for $\F'$ when $n$ is even.

\smallskip

We derive from the above result the proof for $\F'$ when $n=2m-1$ is odd. Recall from Section
\ref{sec:Y_systems}\ref{parg:lax} that the matrices $L'_{2m-1}$ and $B'_{2m-1,2k}$ (for $k=1,\dots,m-1$) were
obtained by slightly modifying the matrices $L'_{2m}$ and $B'_{2m,2k}$ from the previous case. We analyse how these
modifications affect on the one hand the vector field and on the other hand the commutator, which are the left and
right hand sides of the Lax equation
\begin{equation}\label{eq:Lax_odd_even}
  \X_{z_{2k}}L'_{2m-1}=[L'_{2m-1},B'_{2m-1,2k}]\;.
\end{equation}
Before doing this, notice that since by definition the last rows of $L'_{2m-1}$ and of $B'_{2m-1,2k}$ are zero, the
last row of both sides in (\ref{eq:Lax_odd_even}) is zero. Similarly, there is a single non-zero entry in the last
column of $L'_{2m-1}$, and none in the last column of $B'_{2m-1,2k}$, so one only needs to check that
$\pb{x_{2m-1},z_{2k}}_{2m-1}=-z_{2k-1}^{(2m-3)}$. On the one hand, $\pb{x_{2m-1},z_{2k}}_{2m-1}=-\partial
z_{2k}^{(2m-1)}/\partial x_{2m-2}$; on the other hand, it follows from the recursion relation (\ref{eq:z_rec}) that
$z_{2k}^{(2m-1)}=z_{2k}^{(2m-3)}+x_{2m-2}z_{2k-1}^{(2m-3)}$. Next, notice that the remaining entries of the
second-to-last rows and columns of both sides of (\ref{eq:Lax_odd_even}) are zero, since they are already zero in
$L'_{2m}$ and are by definition zero in $B'_{2m-1,2k}$ (except for the diagonal entry). For the other entries
$(i,j)$ of the matrices we compare the vector fields. The recursion relation (\ref{eq:z_rec}) yields
$z_{2k}^{(2m)}=z_{2k}^{(2m-1)}+x_{2m}z_{2k-1}^{(2m-1)}$, which implies that
\begin{equation*}
  \pb{x_i,z_{2k}^{(2m-1)}}_{2m-1}=\left({\pb{x_i,z_{2k}^{(2m)}}_{2m}}\right)_{|_{x_{2m}=0}}\;,
\end{equation*}%
since $i<2m-1$. This means that, except for its two last rows and columns, the matrix $\X_{z_{2k}}L'_{2m-1}$ is
obtained from $\X_{z_{2k}}L'_{2m}$ by substituting $0$ for $x_{2m}$. We need to check that this is also so for the
corresponding submatrix of $[L'_{2m-1},B'_{2m-1,2k}]$. Let $1\leqs i,j\leqs 2m-2$. Then
$(L'_{2m-1})_{i,\ell}(B'_{2m-1,2k})_{\ell,j}=((L'_{2m})_{i,\ell}(B'_{2m,2k})_{\ell,j})_{|_{x_{2m}=0}}$ and
$(B'_{2m-1,2k})_{i,\ell}(L'_{2m-1})_{\ell,j}=(B'_{2m,2k})_{i,\ell}((L'_{2m})_{\ell,j})_{|_{x_{2m}=0}}$ for
$\ell=1,\dots,2m$: for $\ell=1,\dots,2m-2$ this is true by definition, while for $\ell=2m-1$ and $\ell=m$ both
sides are zero.

%

\end{document}